\documentclass[a4paper,10pt]{amsart}
\usepackage{amsmath}
\usepackage[T1]{fontenc}
\usepackage{amssymb}
\usepackage{amsthm}
\newcommand{\md}{{\rm d}}

\newtheorem{thm}{Theorem}[section]
\newtheorem{prop}[thm]{Proposition}
\newtheorem{lem}[thm]{Lemma}

\newtheorem{rem}[thm]{Remark}

\numberwithin{equation}{section}

\title[Generalized Ginzburg-Landau equation]
{On the perfect superconducting  solution for a generalized
Ginzburg-Landau equation}

\author[Ayman Kachmar]{Ayman Kachmar*}
\thanks{$^*$Universit\'e Paris-Sud, D\'epartement de math\'ematiques,
  B\^at. 425, 91405 Orsay France.\break E-mail~:
  ayman.kachmar@math.u-psud.fr} 

\date{\today}

\begin{document}
\maketitle

\begin{abstract}
We study a generalized Ginzburg-Landau equation that models a sample
formed of a superconducting/normal junction and which is not
submitted to an applied magnetic field. We prove the existence of a
unique positive (and bounded)  solution of  this equation. In the
particular case when the domain is the entire plane, we determine
the explicit expression of  the solution (and we find that
it satisfies a Robin  (de\,Gennes) boundary condition on the
boundary of the superconducting side). Using the result of the entire
plane, we determine for the case of general domains, the asymptotic
behavior of the solution for large values of the Ginzburg-Landau
parameter. The main tools are Hopf's Lemma, the Strong Maximum
Principle, elliptic estimates and Agmon type estimates.
\end{abstract}
%\tableofcontents
%\newpage
\section{Introduction and main results}
Let us consider two open, bounded and smooth domains $
\Omega_1,\Omega\subset\mathbb R^2$ such that~:
\begin{equation}\label{VI-Omd}
\overline\Omega_1\subset\Omega,\end{equation}
and let
\begin{equation}\label{VI-Omti}
\Omega_2=\Omega\setminus\overline\Omega_1.
\end{equation}
The domain $\Omega_1$ corresponds to the $2$-D cross section of a
cylindrical superconducting sample with infinite height, and
$\Omega_2$ corresponds to that of a normal material. In the Ginzburg-Landau
theory~\cite{GL}, the superconducting properties are described by a
complex valued wave function $\psi$, called the `order parameter',
whose modulus $|\psi|^2$ measures the density of the superconducting
electron Cooper pairs (hence $\psi\equiv0$ corresponds to the so
called  normal
state), and a real vector field $ A=(A_1,A_2)$, called the `magnetic
potential', such that the induced magnetic field in the
sample corresponds to ${\rm curl}\, A$.  It is well known  (see \cite{Pa, deGe, deGe1}) that when a normal
material is placed adjacent to a superconductor, the superconducting
Cooper electron pairs can diffuse from the superconducting to the
normal material. We then have  to consider pairs $(\psi,A)$ defined on $\Omega$.\\
The basic postulate in the Ginzburg-Landau theory is that the pair
$(\psi,A)$ minimizes the Gibbs free energy, which, in our case, has  the following
dimensionless form~\cite{Chetal}~:
\begin{eqnarray}\label{VI-EGL}
&&\\
\mathcal G(\psi,A)&=&\int_{\Omega_1}
\left\{|(\nabla-iA)\psi|^2+\frac1{2\varepsilon^2}(1-|\psi|^2)^2+|{\rm
curl}\,A-H|^2\right\}\md x\nonumber\\
&&+\int_{\Omega_2}\left\{\frac1m|(\nabla-iA)\psi|^2+\frac{a}{\varepsilon^2}|\psi|^2+
\mu\left|\frac1\mu{\rm curl}\,A-H\right|^2\right\}\md x.\nonumber
\end{eqnarray}
Here,
$\frac1\varepsilon$ is the Ginzburg-Landau parameter, a
characteristic of the superconducting material (filling ${\Omega_1}$),
$m>0$ is a characteristic of the normal material (filling
$\Omega_2$), $\mu>0$ is the magnetic permeability in $\Omega_2$, $H>0$
is the intensity of the applied magnetic field and
$a>0$ is related to the critical temperature of the material in
$\Omega_2$. The positive sign of $a$ means that we are above the
critical temperature of the normal material.\\
The procedure of modeling normal materials by taking a positive sign
of the quadratic term $|\psi|^2$ in the Ginzburg-Landau energy 
has been the subject of a vast mathematical
literature. We do not aim at citing a complete list but we refer to
\cite{AlBr, DuRe, Gi, RuSt1, RuSt} and the references therein.\\ 
%The functional (\ref{VI-EGL}) is gauge invariant, i.e. given $\chi\in
%H^2(\Omega)$, we have,
%$$\mathcal G(\psi,A)=\mathcal G(\psi\exp(i\chi),A+\nabla\chi).$$
%It turns out (cf.~\cite{BR}) that the
%space $\mathcal H=H^1(\Omega;\mathbb C)\times
%H^1(\Omega;\mathbb R^2)$ is that of finite energy
%configurations for (\ref{VI-EGL}). 
Minimization of  the
functional (\ref{VI-EGL}) will take place in  the space $$\mathcal H=H^1(\Omega;\mathbb C)\times
H^1(\Omega;\mathbb R^2).$$ The functional (\ref{VI-EGL}) is gauge
invariant in the sense that  given $\chi\in H^2(\Omega)$, we have,
$$\mathcal G(\psi,A)=\mathcal G(\psi\,e^{i\chi},A+\nabla\chi).$$
When the applied magnetic field $H=0$,
we shall see that the  minimizers of (\ref{VI-EGL}) are
completely determined by those of the functional (which is
naturally obtained by taking $A=0$
and $H=0$ in (\ref{VI-EGL}))~:
\begin{equation}\label{VI-EnGL1}
\mathcal G_0(u)=\int_{{\Omega_1}}\left(|\nabla
u|^2+\frac1{2\varepsilon^2}(1-|u|^2)^2\right)\md
x+\int_{\Omega_2}\left(\frac1m|\nabla
u|^2+\frac{a}{\varepsilon^2}|u|^2\right)\md x.
\end{equation}
We emphasize that the functional (\ref{VI-EnGL1}) is defined for
real-valued functions in $H^1(\Omega;\mathbb R)$.
We shall show that the minimizers 
of the functional (\ref{VI-EnGL1}) are completely 
determined by the positive solutions  of the following
`generalized' Ginzburg-Landau equation
(see Theorem~\ref{VI-LuPaTh1} below)~:
\begin{equation}\label{VIGL}
\left\{\begin{array}{l}
-\Delta u=\displaystyle\frac1{\varepsilon^2}(1-|u|^2)u,\quad \text{in }\Omega_1,\\
\\
-\displaystyle \frac1m\Delta u=-\frac{1}{\varepsilon^2}a\,u,\quad \text{in
}\Omega_2,\\
\\
\mathcal T_{\partial\Omega_1}^{\rm int}\left(\displaystyle\frac{\partial
u}{\partial\nu_1}\right)=\displaystyle\frac1m\mathcal
T_{\partial\Omega_1}^{\rm ext}\left(\displaystyle\frac{\partial u}{\partial\nu_1}\right),\quad
\text{on }\partial\Omega_1,\\
\\
\mathcal T_{\partial\Omega}^{\rm int}\left(\displaystyle\frac{\partial
u}{\partial\nu}\right)=0\quad \text{on
}\partial\Omega.\end{array}\right.
\end{equation}
Here, $\nu_1$ is the unit
outer normal of the boundary $\partial\Omega_1$, $\nu$ that of
$\partial\Omega$, and, given an open set $U\subset\mathbb
R^2$,  $\mathcal T_{\partial U}^{\rm int}$ and $\mathcal T_{\partial
U}^{\rm ext}$ denote respectively the interior and exterior trace
operators on $\partial U$~:
$$\mathcal T_{\partial U}^{\rm int}:H^1(U)\longrightarrow L^2(\partial U),\quad
\mathcal T_{\partial U}^{\rm ext}:H^1_{\rm loc}(U^c)\longrightarrow
L^2(\partial U).$$
%%%%%%%%%%%%%%%%%%%%%%%%%%%%%%%%%%%%%%%%weakformulation%%%%%%%%%%%%%%%%%%%%%%%%%%%%%%%%%%%%%%%%%%%%%
%We have the following  weak formulation of
%equation (\ref{VIGL})~: A function $u\in H^1(\Omega)$ is
%said to be a weak solution of (\ref{VIGL}) if the following
%condition holds,
%\begin{eqnarray*}
%&&\int_{\Omega_1}\nabla u\cdot\nabla v\,\md
%x+\frac1m\int_{\Omega_2}\nabla
%u\cdot\nabla v\,\md x\\
%&&=\frac1{\varepsilon^2}\left( \int_{\Omega_1}(1-|u|^2)u\,v\,\md
%x-a\int_{\Omega_2}u\,v\,\md x\right),\quad\forall v\in
%H^1(\Omega).
%\end{eqnarray*}
The existence, uniqueness and asymptotic behavior (as
$\varepsilon\to0$) of the non-negative
solutions of equation (\ref{VIGL}) will be the main concerns of this
paper.
We believe that a careful understanding of this situation  will be
useful for the investigation of
the behaviour of minimizers of (\ref{VI-EGL}) as the applied magnetic
field increases from $H=0$ (this will be hopefully the subject of a
forthcoming work). Actually, physicists (cf. \cite{Pa})
claim that minimizers of (\ref{VI-EGL}) are very sensitive to the
variations of the applied magnetic field $H$, even when it remains  small.\\
Given $a,m,\varepsilon>0$, we define the following eigenvalue~:
\begin{eqnarray}\label{VI-EV}
&&\lambda_1(a,m,\varepsilon)=\inf\left\{\int_{\Omega_1}\left(|\nabla\phi|^2
-\frac1{\varepsilon^2}|\phi|^2\right)\md x\right.\\
&&\left.+\int_{\Omega_2}\left(
\frac1m|\nabla\phi|^2+\frac{a}{\varepsilon^2}|\phi|^2\right)\md
x~:\quad \phi\in
H^1(\Omega),\,\|\phi\|_{L^2(\Omega)}=1\right\}.\nonumber
\end{eqnarray}

In the theorem below, we establish
the relation between the minimizers of
(\ref{VI-EGL}) and the positive solution of (\ref{VIGL}).

\begin{thm}\label{VI-LuPaTh1}
With the previous notations, the
following assertions hold.
\begin{enumerate}
\item If $\lambda_1(a,m,\varepsilon)< 0$, 
then (\ref{VIGL}) admits  a non-negative non-trivial solution. If, furthermore
$\partial\Omega_1,\partial\Omega$ are of class $C^3$, then this
solution  is unique  and satisfies
$0<u_\varepsilon<1$ in $\overline\Omega$.
\item If $\lambda_1(a,m,\varepsilon)\geq0$, then the unique solution
of (\ref{VIGL}) is the trivial solution $u_\varepsilon\equiv0$.
\item
If the applied magnetic field $H=0$ and $(\psi,A)$ is a minimizer
of (\ref{VI-EGL}), then $|\psi|\equiv u_\varepsilon$.
\item If $\Omega$ is simply connected and if $H=0$, then
the set of minimizers of (\ref{VI-EGL}) in $H^1(\Omega;\mathbb
C)\times H^1(\Omega;\mathbb R^2)$
 is given by,
$$\{(u_\varepsilon e^{i\chi},\nabla\chi)~:\quad \chi\in
H^2(\Omega)\}.$$
\end{enumerate}
\end{thm}

Notice that if  $\varepsilon\in]0,1/{\sqrt{\lambda_1(\Omega_1)}}[$
(here, given a bounded regular open set $\mathcal O\subset\mathbb R^2$,
$\lambda_1(\mathcal O)$ 
denotes the first eigenvalue  of the Dirichlet realization of
$-\Delta$ in $\mathcal O$), then $\lambda_1(a,m,\varepsilon)<0$. This follows
directly from   the min-max principle, which gives~:
\begin{equation}\label{VI-Min-Max}
\lambda_1(a,m,\varepsilon)\leq
\min\left(\lambda_1({\Omega_1})-\frac1{\varepsilon^2},\frac1m\lambda_1(\Omega_2)
+\frac{a}{\varepsilon^2}\right).
\end{equation}
Hence, in this case, the solution
$u_\varepsilon$ of Theorem~\ref{VI-LuPaTh1} is non-trivial, and
we shall investigate, in Theorem~\ref{VI-LuPaTh2}, its
asymptotic behavior as
$\varepsilon\to0$.\\
We define the function $\mathbb R\ni t\mapsto U(t)$ by~:
\begin{equation}\label{VI-Sol}
U(t)=\left\{
\begin{array}{l}
\displaystyle\frac{\beta\exp({\sqrt{2}\,t})-1}{\beta\exp({\sqrt{2}\,t)+1}},\quad
t\geq0,\\
\\
A\,\exp(\sqrt{am}\,t),\quad t<0,
\end{array}\right.
\end{equation}
where the constants $\beta$ and $A$ are given by~:
\begin{equation}\label{VI-l,bet}
\beta=\frac{\sqrt{2m}+\sqrt{a+2m}}{\sqrt{a}},\quad
A=\frac{\sqrt{2m}+\sqrt{a+2m}-\sqrt{a}}{\sqrt{2m}+\sqrt{a+2m}+\sqrt{a}}.
\end{equation}

\begin{thm}\label{VI-LuPaTh2}
Let $\varepsilon\in]0,\frac1{\sqrt{\lambda_1({\Omega_1})}}[$ and
$u_\varepsilon$ be the unique positive solution of (\ref{VIGL}).
Then, for any compact sets $K_1\subset\Omega_1$, $K_2\subset
\Omega_2$,  we have as  $\varepsilon\to0$,
\begin{equation}\label{VIu-lim0}
u_\varepsilon\to1\quad\text{in }C^2({K_1}),\quad
u_\varepsilon\to0\quad\text{in }C^2(K_2).
\end{equation}
Moreover, if $\partial\Omega_1,\partial\Omega$ are of class $C^4$, then there exists a function $w_\varepsilon\in
C(\overline\Omega)$ that converges to $0$ uniformly in
$\overline\Omega$ and such that
\begin{equation}\label{VI-bndlay-S}
u_\varepsilon(x)=U\left(\frac{t(x)}\varepsilon\right)+w_\varepsilon(x),\quad\forall~x\in\overline{\Omega}.
\end{equation}
Here, the function $U$ is defined by (\ref{VI-Sol}) 
and the function $t$ is defined by
\begin{equation}\label{VI-t}
t(x)=\left\{\begin{array}{l}
{\rm dist}(x,\partial{\Omega_1}),\quad x\in{\Omega_1},\\
-{\rm dist}(x,\partial{\Omega_1}),\quad x\in\mathbb
R^2\setminus{\Omega_1}.
\end{array}\right.
\end{equation}
\end{thm}

\begin{rem}\label{VIcomm}
Theorem~\ref{VI-LuPaTh2} shows that the solution $u_\varepsilon$
exhibits a boundary layer near $\partial{\Omega_1}$ with scale $\mathcal
O(\varepsilon)$ as $\varepsilon\to0$. Remembering the physical
interpretation\footnote{ $u_\varepsilon^2$ measures the density of
the superconducting Cooper electron pairs.} of $u_\varepsilon$, we
see that the thickness of the superconducting region in $\Omega_2$  is
$\mathcal
O(\varepsilon)$.
\end{rem}

In the next theorem, we give an asymptotic expansion  of the energy
(\ref{VI-EnGL1}) of the
positive solution $u_\varepsilon$ as $\varepsilon\to0$.

\begin{thm}\label{VIen}
Under the hypotheses of Theorem~\ref{VI-LuPaTh2}, 
the following asymptotic expansion holds~:
\begin{equation}\label{VIen-eq}
\mathcal G_0(u_\varepsilon)
=
%\left(\frac{4\sqrt{2}(3\beta+1)}{3(\beta+1)^3}
%+\frac12\sqrt{\frac am}\left(1+\frac1{m}\right)A^2
%+o(1)\right)
c_1(a,m)\frac{|\partial{\Omega_1}|}\varepsilon-c_2(a,m)
\int_{\partial\Omega_1}\kappa_{\rm r}(s)\,\md s+o(1),
\quad(\varepsilon\to0).
\end{equation}
Here $c_1(a,m)$  and $c_2(a,m)$ are two positive parameters
and $\kappa_{\rm r}$ is the scalar curvature of the boundary of $\Omega_1$.
\end{thm}

The expressions of the constants $c_1(a,m)$ and $c_2(a,m)$ will be
given in (\ref{VI-c1+}) and (\ref{VI-c2+}) respectively.\\

The asymptotic behavior of the solution $u_\varepsilon$ is based on
the understanding of the `model' equation associated to $\Omega_1=\mathbb
R\times\mathbb R_+$ and $\Omega_2=\mathbb R\times\mathbb R_-$. Due to
the invariance by scaling of $\mathbb R\times\mathbb R_\pm$, we are
reduced in this case to the following equation (i.e. with $\varepsilon=1$)~:
\begin{equation}\label{VI-limeq}
\left\{\begin{array}{l} -\Delta
u=(1-u^2)u,\quad\text{in }\mathbb
R\times\mathbb R_+,\\
\\
-\displaystyle\frac1m\Delta u=-a\,u,\quad\text{in }\mathbb
R\times\mathbb R_-,\\
\\
\left(\displaystyle\frac{\partial u}{\partial
x_2}\right)(x_1,0_+)=\displaystyle\frac1m\left(\frac{\partial u}{\partial
x_2}\right)(x_1,0_-),\quad u(x_1,0_+)=u(x_1,0_-),\quad\text{on
}\mathbb R.
\end{array}\right.
\end{equation}
Notice that the function $(x_1,x_2)\mapsto U(x_2)$, where $U$ is
defined by (\ref{VI-Sol}), is a solution of (\ref{VI-limeq}).\\
Since Equation (\ref{VI-limeq}) arises as a limiting form of
(\ref{VIGL}), we focus on solutions of  (\ref{VI-limeq}) that are in the
class
\begin{equation}\label{VI-C}
\mathcal C=\{u\in  L^\infty(\mathbb
  R^2)~:~u_{|_{\mathbb R\times\mathbb R_\pm}}\in C^2(\overline{\mathbb
  R\times\mathbb R_\pm}),\quad
~u\geq0 \text{ in }\mathbb R^2\}.\end{equation}

\begin{thm}\label{VI-sol-modeq}
Equation (\ref{VI-limeq}) admits a unique solution in 
$\mathcal C$, which is given by~:
$$\mathbb R^2\ni (x_1,x_2)\mapsto U(x_2),$$
where $U$ is the function defined by (\ref{VI-Sol}).
\end{thm}

Notice that the solution $U$ of (\ref{VI-limeq}) satisfies the
following boundary condition on the boundary of $\mathbb
R\times\mathbb R_+$~:
\begin{equation}\label{VI-deGebndcn}
\frac{\partial U}{\partial x_2}(0_+)=\gamma\,U(0_+),
\end{equation}
where $\gamma$ is given by~:
$$\gamma=\sqrt{\frac am}.$$
This `Robin' boundary condition was already present in the physics literature (cf.~\cite{deGe}),
and it is called in that context
`de\,Gennes boundary condition'. The intuitive reason for deriving
this boundary condition from (\ref{VI-limeq}) is that the equation in
$\mathbb R\times\mathbb R_+$ is linear and its solution in $\mathcal
C$ is a simple exponential function
$A\,\exp\left(\sqrt{am}\,x_2\right)$. Then we get the boundary condition on
$\mathbb R\times\mathbb R_+$ from that in (\ref{VI-limeq}), see
Section~4 for more details.\\ 
%\begin{rem}\label{VIen}
%Theorem~\ref{VI-LuPaTh2} shows that the energy associated with
%$u_\varepsilon$ is of the order $\mathcal O(\varepsilon^{-1})$,
%$$\mathcal G_0(u_\varepsilon)=\mathcal O\left(\frac1\varepsilon\right),\quad
%(\varepsilon\to0).$$ Here the energy functional  $\mathcal G_0$ is
%defined in (\ref{Vi-EnGl1}).
%\end{rem}

In \cite{LuPa96}, the authors study the following Ginzburg-Landau
equation with `de\,Gennes boundary condition'~:
\begin{equation}\label{VI-GL-LuPa}
\left\{\begin{array}{l} -\Delta
u=\displaystyle\frac1{\varepsilon^2}(1-u^2)u,\quad
\text{in }{\Omega_1},\\
-\displaystyle\frac{\partial u}{\partial\nu_1}=\gamma(\varepsilon)u,\quad \text{on
}\partial{\Omega_1},\end{array}\right.
\end{equation}
where $\gamma(\varepsilon)\geq0$ is `the de\,Gennes parameter' that
may depend on $\varepsilon$.\\
In the case when  $\gamma(\varepsilon)=0$ (which corresponds
to the situation when the superconductor is  adjacent to a
vacuum),  it is well known that $u\equiv\pm1$ are  the only solutions of
(\ref{VI-GL-LuPa}) (see e.g. \cite{CHO, DGP}).
These solutions reveal  perfect superconducting
states. Compared with
our results (Theorems~\ref{VI-LuPaTh1} and~\ref{VI-LuPaTh2}), we
observe that the presence of a normal material exterior to a
superconductor has a strong effect on the perfect superconducting
solution. This complements the picture initiated in our previous
work~\cite{kach2} (see also~\cite{GJ}), where we showed that the
presence of a normal material adjacent to a superconductor can also have  a
strong influence on the
onset of superconductivity.\\
Lu and Pan~\cite{LuPa96} study the asymptotic behavior of the
positive solution of (\ref{VI-GL-LuPa}) when $\gamma(\varepsilon)>0$
and as $\varepsilon\to0$. Just as in our case, they obtained  that
the case of $\gamma(\varepsilon)\not=0$ is quite different from the
case of $\gamma(\varepsilon)=0$ (cf.~\cite[Theorem~2]{LuPa96}). In
particular, they obtained that if
$$0<\lim_{\varepsilon\to0} \varepsilon\gamma(\varepsilon)<+\infty,$$
then the positive solution of (\ref{VI-GL-LuPa}) exhibits a boundary
layer and shows a similar behavior to that of equation (\ref{VIGL})
(cf. Theorem~\ref{VI-LuPaTh2}). We also point out that they  give only
the leading order term of the energy of the solution, whereas we
obtain a two-term expansion of our minimizing energy.\\

In the presence of  an applied magnetic field $H>0$, the situation  is
more or less related  to the phenomenon of pinning
(cf. e.g. \cite{AfSaSe}). Pinning models replace the usual potential
term in the Ginzburg-Landau energy functional by 
$(a_\varepsilon-|\psi|^2)^2$, where $a_\varepsilon$, {\it the maximal
  superconducting density}, is   a smooth function. If one has to
recover our case, the function $a_\varepsilon$ would be a step
function, equal to $1$ in $\Omega_1$, and equal to $-a<0$ in $\Omega_2$.\\
Let us mention
that a standard application of the maximum principle shows that if
$(\psi,A)$ is a
minimizer  of (\ref{VI-EGL}), then $|\psi|\leq u_\varepsilon$.
Coming back to the asymptotic profile of $u_\varepsilon$, we notice
that it satisfies (cf. Theorem~\ref{VI-LuPaTh2} and more precisely
Formula (5.20))
$$|\nabla u_\varepsilon|\geq \frac{C}{\varepsilon}\quad{\rm in~a~
neighborhood~of~}\partial\Omega_1.$$
The authors of \cite{AfSaSe} consider a non-constant  maximal superconducting
density $a_\varepsilon$ but with the restriction that it can not  
oscillate quicker than $|\ln\varepsilon|$ ($|\nabla a_\varepsilon|\leq
C|\ln\varepsilon|$). One other complication for the case with magnetic
field comes from the structure
of the functional (\ref{VI-EGL}) where the term $(1-|\psi|^2)^2$ is absent
from the integrand in $\Omega_2$, hence one can no more obtain the
localization of the
`vortex-balls' in $\Omega_2$ by applying directly the co-area formula
as was done in \cite{SaSe00}.\\

Perhaps it is  the Ginzburg-Landau equation with  Dirichlet boundary condition that has
received the early attention in the literature (cf.~\cite{BBH, L, S}).
Actually, the solution of the following Dirichlet problem
\begin{equation}\label{VI-GL-D}
\left\{\begin{array}{l} -\Delta
u=\displaystyle\frac1{\varepsilon^2}(1-|u|^2)u,\quad \text{in }{\Omega_1},\\
u=g,\quad\text{on }\partial{\Omega_1},
\end{array}\right.
\end{equation}
where $g$ is a complex-valued mapping from $\partial{\Omega_1}$ to the
unit circle $\mathbb S^1$, can exhibit a vortex structure (depending
on the Brouwer degree of $g$).
%and its associated
%energy is of the order $\mathcal O(\log\varepsilon)$ as
%$\varepsilon\to0$.
This shows that this situation  is quite different
from ours.\\

%%%%%%%%%%%%%%%%%%%%%energy estimate%%%%%%%%%%%%%%%%%%%%%%%%%%%%%%%%%%%%%%%%%%%%
% where the energy of the
%minimal solution of (\ref{VIGL}) is of the order $\mathcal
%O(\varepsilon^{-1})$ (cf. Remark~\ref{VIen}).\\
%%%%%%%%%%%%%%%BOUNDARYCONDITION=0%%%%%%%%%%%%%%%%%%%%%%%%%%%%%%%%%%%%%%%%%%%%%%%
%Let us also mention  that (\ref{VI-GL-D}) with $g=0$ could be of
%physical interest and is useful for modeling the case when ${\Omega_1}$
%is adjacent to a ferromagnetic material (cf.~\cite{GP, HTW}).
%However, the analysis of Lu-Pan recovers also this equation and
%permits one to obtain that the energy of a least-energy solution  has
%the order of $\mathcal O(\varepsilon^{-1})$ (cf.~\cite[Theorem~2,
%Case~(3)]{LuPa96}) contrary to the case when
%$|g|=1$.\\%%%%%%%%%%%%%%%%%%%%%%%%%%%%%%%%%%%%%%%%%%%%%%%%%%%%%%%%%%%%%%%%%%%%%%
\noindent We present now the outline of the paper.\\ 
\indent In Section~2, we give some
auxiliary material that we shall use frequently in the paper and we
discuss the regularity of weak solutions to Equation (\ref{VIGL}).
In Section~3,  we prove the existence and uniqueness of the positive
solution to Equation (\ref{VIGL}), and  we finish the proof of
Theorem~\ref{VI-LuPaTh1}.\\
In Section~4, we study the uniqueness of bounded solutions for Equation (\ref{VI-limeq}) and we prove
Theorem~\ref{VI-sol-modeq}.
Using  the result of Theorem~\ref{VI-sol-modeq}, 
we are able to describe in Section~5, by the use of 
elliptic estimates together with an analysis near $\partial{\Omega_1}$,
the asymptotic behavior of the positive solution $u_\varepsilon$ as
$\varepsilon\to0$, proving thus Theorem~\ref{VI-LuPaTh2}. 
In Section~6, we determine  the energy estimate of
Theorem~\ref{VIen} through an auxiliary result concerning a
one-dimensional variational problem. Finally,
we give in Section~7 some concluding remarks and we shed light on
some points that seem to  us interesting for further research.

\section{Preliminaries}\label{VIsec2}

\subsection{A maximum principle}\ \\
When analyzing the behavior of the solution of (\ref{VIGL}), we shall need frequently the following variant of the
maximum principle, which we take from \cite[Lemma~3.4 and Theorem~3.5]{GiTr}.
\begin{thm}\label{VI-MP-GiTr}
Consider an open connected set  $U\subset\mathbb R^2$ having  a
smooth boundary of class $C^1$. 
Let $w\in C^2(U)\cap C^1(\overline
U)$ and $c\in L^\infty(U)$ be bounded functions. Suppose that  $-\Delta w+c(x)w\geq0$,
$c(x)\geq0$ in $U$, and that there exists $x_0\in \overline U$ such that $w(x_0)=\displaystyle\min_{x\in\overline
U}w(x)\leq0$. Then~:
\begin{enumerate}
\item If $w(x)>w(x_0)$ in $\overline U$ and $x_0\in\partial U$, $(\partial w/\partial v)(x_0)<0$;
\item If $x_0\in U$, $w(x)\equiv w(x_0)$.
\end{enumerate}
\end{thm}

Assertion (1) in Theorem~\ref{VI-MP-GiTr} corresponds to `Hopf's Lemma' while Assertion (2) is the `Strong Maximum Principle'.

\subsection{Boundary coordinates}\label{VI-BndCord}
For the analysis of the behavior of the solution of (\ref{VIGL}) near
the boundary $\partial\Omega_1$, we often
write the equation in a coordinate
system valid near $\partial{\Omega_1}$. Suppose that $\partial {\Omega_1}$
is smooth of class $C^{k+2}$, with $k\in\mathbb N$. Given
$t_0>0$, we define the following subset~:
\begin{equation}\label{VI-Om-t}
{\Omega_1}(t_0)=\{x\in\mathbb R^2~:\quad {\rm
dist}(x,\partial{\Omega_1})<t_0\}.
\end{equation}
We define also the function $t~:\mathbb R^2\mapsto\mathbb R$ by,
\begin{equation}\label{VI-t(x)}
t(x)=\left\{\begin{array}{l} {\rm dist}(x,\partial{\Omega_1}),\quad
x\in{\Omega_1},\\
-{\rm dist}(x,\partial{\Omega_1}),\quad x\in\mathbb R^2\setminus{\Omega_1}.
\end{array}\right.
\end{equation}
We can choose $t_0$ sufficiently small so that $t\in
C^{k+2}({\Omega_1}(t_0))$ and $\nabla t(x)=-\nu_1(s(x))$
(cf.~\cite[Section~14.6]{GiTr}). Here $s(x)\in\partial{\Omega_1}$ is the unique point
defined by
$${\rm dist}(x,s(x))={\rm dist}(x,\partial{\Omega_1}),$$
and $\nu_1$ is the unit outward normal of $\partial{\Omega_1}$.\\
Let us consider also a parametrization
$$s\in]-\frac{|\partial{\Omega_1}|}2,\frac{|\partial{\Omega_1}|}2]\mapsto
M(s)\in\partial{\Omega_1}$$ of $\partial{\Omega_1}$ that satisfies~:
$$\left\{
\begin{array}{l}
s \text{ is the oriented `arc length' between }M(0) \text{ and }
M(s);\\
T(s):=M'(s) \text{ is a unit tangent vector
to }\partial{\Omega_1}\text{ at the point }M(s);\\
\text{The orientation is positive, i.e. }{\rm det}(T(s),\nu_1(s))=1.
\end{array}
\right. $$ We recall that $\nu_1(s)$ is the unit outward normal of
$\partial{\Omega_1}$ at the point $M(s)$. The scalar curvature
$\kappa_{\rm r}$ is now defined by~:
\begin{equation}\label{kappa-r}
T'(s)=\kappa_{\rm r}(s)\nu_1(s).
\end{equation}
We define now the following coordinates transformation~:
\begin{equation}\label{Phi(s,t)}
\Phi:\,]-|\partial{\Omega_1}|/2,|\partial{\Omega_1}|/2]\times
]-t_0,t_0[\,\ni(s,t)\mapsto M(s)-t\nu_1(s)\in {\Omega_1}(t_0).
\end{equation}
Then $\Phi$ is a $C^{k+1}$-diffeomorphism, and for
$x\in{\Omega_1}(t_0)$, we write,
\begin{equation}\label{Phi-1}
\Phi^{-1}(x):=(s(x),t(x)).
\end{equation}
The Jacobian of the transformation $\Phi^{-1}$ is given by,
\begin{equation}\label{Jac}
{\rm a}(s,t)={\rm det}\left(D\Phi^{-1}\right)=1-t\kappa_{\rm r}(s).
\end{equation}
For a function $u\in H^1_0({\Omega_1}(t_0))$, we define a function
$\widetilde u\in H^1(\Phi^{-1}({\Omega_1}(t_0)))$ by~:
\begin{equation}\label{VI-Tu}
\widetilde u(s,t)= u(\Phi(s,t)). \end{equation} Then we have the
following change of variable formulas,
\begin{equation}\label{VI-NL2}
\int_{{\Omega_1}(t_0)}|u(x)|^2\md
x=\int_{-|\partial{\Omega_1}|/2}^{|\partial{\Omega_1}|/2}\int_{-t_0}^{t_0}|\widetilde
u(s,t)|^2 {\rm a}(s,t)\,\md s\md t, \end{equation} and, for any function
$v\in H^1_0({\Omega_1}(t_0))$,
\begin{equation}\label{VI-WL2}
\int_{{\Omega_1}(t_0)}\nabla u(x)\cdot\nabla v(x)\,\md
x=\int_{-|\partial{\Omega_1}|/2}^{|\partial{\Omega_1}|/2}\int_{-t_0}^{t_0}\left\{(\partial_t\widetilde
u)(\partial_t\widetilde v)+{\rm a}^{-2}(\partial_s\widetilde
u)(\partial_s\widetilde v)\right\}{\rm a}(s,t)\,\md s\md t.
\end{equation}
This last formula permits us to write (in the sense of
distributions)~:
\begin{equation}\label{VI-TLap}
\left(\Delta u\right)(x)=\left(\widetilde\Delta\, \widetilde
u\right)(\Phi^{-1}(x)), \quad \text{in }\mathcal D'( {\Omega_1}(t_0)),
\end{equation}
where the differential operator $\widetilde\Delta$ is defined by~:
\begin{equation}\label{VI-TDel}
\widetilde\Delta= {\rm a}^{-2}(s,t)\frac{\partial^2}{\partial
s^2}+\frac{\partial^2}{\partial t^2}+\left(t\kappa'_{\rm
r}(s){\rm a}^{-3}(s,t)\right)\frac{\partial}{\partial s}-\left(\kappa_{\rm
r}(s){\rm a}^{-1}(s,t)\right)\frac{\partial}{\partial t}.
\end{equation}
%%%%%%%%%%%%%%%%%%%%%%%%%%%%%%%%%%%%%estmateofH^knorms%%%%%%%%%%%%%%%%%%%%%%%%%%%%%%%%%%%%%%%%%%%%%%%%%%%%%%%%%%%%%%
%We get now the following lemma by a straight-forward calculation.
%
%\begin{lem}\label{VI-Lem-com}
%Let $\Delta'$ be the Laplacian operator in $(s,t)$-coordinates,
%\begin{equation}\label{VI-Ls,t}
%\Delta'=\frac{\partial^2}{\partial s^2}+\frac{\partial^2}{\partial
%t^2}.
%\end{equation}
%Suppose that
%$$u\in H^{k+2}_{\rm loc}({\Omega_1}),\quad {\rm supp}\,u\subset{\Omega_1}(t_0)_{_+},$$
%where $k\in\{0,1,2,\cdots\}$ and
%${\Omega_1}(t_0)_{_+}={\Omega_1}(t_0)\cap\overline{{\Omega_1}}$. Then there
%exists a constant $C>0$ depending only on ${\Omega_1},\Omega_2$ and $k$ such
%that,
%\begin{equation}\label{VI-estLL'}
%\|\Delta'\, \widetilde u\|_{H^k(I_{_+})}\leq \|\widetilde\Delta\,
%\widetilde u\|_{H^k(I_{_+})}+C\left\{t_0\|\widetilde
%u\|_{H^{k+2}(I_{_+})}+\| \widetilde
%u\|_{H^{k+1}(I_{_+})}\right\},\nonumber
%\end{equation}
%and
%\begin{equation}\label{VI-estLL'2}
%\|\widetilde\Delta\, \widetilde u\|_{H^{k+2}(I_{_+})}\leq
%\|\Delta'\, \widetilde u\|_{H^k(I_{_+})}+C\left\{t_0\|\widetilde
%u\|_{H^{k+2}(I_{_+})}+\| \widetilde
%u\|_{H^{k+1}(I_{_+})}\right\},\nonumber
%\end{equation}
%where $I_{_+}=\Phi^{-1}({\Omega_1}(t_0)_{_+})$ and the function
%$\widetilde u$ is associated to $u$ by means of change of variables
%(cf. (\ref{VI-Tu})).
%\end{lem}

%\begin{rem}
%A similar statement to that of Lemma~\ref{VI-Lem-com} holds if the
%support of the function $u$ is in ${\Omega_1}(t_0)_{_-}$, where
%${\Omega_1}(t_0)_{_-}={\Omega_1}(t_0)\cap \overline{\Omega_2}$.
%\end{rem}

\subsection{A regularity result}\label{VIsubsec2.1}\label{VIsubsec2.2}
In this  section we state a regularity theorem adapted to solutions  of (\ref{VIGL}).

\begin{thm}\label{VI-regKa}
Suppose that ${\Omega_1}\subset\mathbb R^2$ has a compact boundary of
class $C^{k+2}$, with $k\in\mathbb N$. There exists a constant $C_k>0$
such that if  $u\in
H^1_0(\Omega)$ and $f\in L^2(\Omega)$ satisfy~:
\begin{equation}\label{VI-WF-Delta=f}
\forall v\in H^1_0(\Omega),\quad\int_{\Omega_1}\nabla u\cdot\nabla
v\,\md x+\frac1m\int_{\Omega_2}\nabla u\cdot\nabla v\,\md
x=\int_\Omega f\,v\,\md x,
\end{equation}
$$f_{|_{\Omega_1}}\in H^k({\Omega_1}),\quad f_{|_{\Omega_2}}\in H^k(\Omega_2),$$
then
$$u_{|_{{\Omega_1}}}\in H^{k+2}({\Omega_1}),\quad u_{|_{\Omega_2}}\in
H^{k+2}(\Omega_2),$$ and we have the following estimate,
\begin{eqnarray*}
&&\|u\|_{H^{k+2}({\Omega_1})}+\|u\|_{H^{k+2}(\Omega_2)}\\
&&\leq C_k\left(\|f\|_{H^k({\Omega_1})}+\|f\|_{H^k(\Omega_2)}+\|u\|_{L^2(\Omega)}\right).
\end{eqnarray*}
\end{thm}

To our knowledge, Theorem~\ref{VI-regKa} is not present
in the former literature.
The proof of Theorem~\ref{VI-regKa} involves the technique of difference quotients (see
\cite{LiMa}), and is given  in
Appendix~B.

\section{Existence and uniqueness in bounded domains}
Let us consider the functional $\mathcal G_0$ introduced in
(\ref{VI-EnGL1}). We denote its
minimum over $H^1(\Omega;\mathbb R)$ by~:
\begin{equation}\label{VI-infEn}
C_0(\varepsilon):=\inf_{u\in H^1(\Omega;\mathbb R)}\mathcal G_0(u).
\end{equation}
It is standard, by starting from a minimizing sequence, to prove the
existence of minimizers of the functional $\mathcal G_0$. Notice
also that minimizers of $\mathcal G_0$ are weak solutions of
Equation (\ref{VIGL}). In all what follows we shall always write
$H^1(\Omega)$ for $H^1(\Omega;\mathbb R)$ and we emphasize that
minimization of the functional $\mathcal G_0$ will be always
considered over real-valued $H^1$ functions.\\

We shall need the following lemma.

\begin{lem}\label{VI-lemSec3}
Let $u$ be a weak solution of (\ref{VIGL}) such that $u\not\equiv0$.
Then,
\begin{enumerate}
%\item $|u(x)|\leq1\quad \forall x\in\overline{\Omega}$;
\item $u\not\equiv0$ in ${\Omega_1}$;
\item $u\not\equiv0$ in $\Omega_2$.
\end{enumerate}
\end{lem}
\paragraph{\bf Proof.}
%Assertion (1) is standard, following the argument of~\cite{DGP} for
%example.\\
We prove assertion (1). Suppose by contradiction that $u\equiv0$
in ${\Omega_1}$. Then, using the transmission property (i.e. the
boundary condition on the interior boundary $\partial\Omega_1$, cf.
(\ref{VIGL})),
$-ma<0$ will be an eigenvalue of the Neumann
Laplacian $-\Delta$ in $\Omega_2$, which is impossible.\\
We prove assertion (2). If $u\equiv0$ in $\Omega_2$, then $u$
satisfies in ${\Omega_1}$,
$$-\Delta u=\frac1{\varepsilon^2}(1-u^2)u,$$
with Neumann boundary condition $\partial u/\partial\nu_1=0$ on
$\partial{\Omega_1}$. Then, by~\cite{DGP}, $|u|\equiv1$ in ${\Omega_1}$ which
contradicts the fact that $u\in H^1(\Omega)$.
\hfill$\Box$\\

 Let us recall the definition of the eigenvalue
$\lambda_1(a,m,\varepsilon)$ given in (\ref{VI-EV}). In the next
proposition, we determine, through  the sign of
$\lambda_1(a,m,\varepsilon)$, the regime of $a,m,\varepsilon$ for which a non-zero solution
of (\ref{VIGL}) exists.

\begin{prop}\label{VIVI-LuPaP2.1}
If $\lambda_1(a,m,\varepsilon)\geq0$, then (\ref{VIGL}) has as a
unique solution $u\equiv0$, which is the unique minimizer of (\ref{VI-EnGL1}).\\
In addition, if $\lambda_1(a,m,\varepsilon)<0$, then $u\equiv0$ is not
a minimizer of (\ref{VI-EnGL1}).
\end{prop}
\paragraph{\bf Proof.}
Let us suppose that $\lambda_1(a,m,\varepsilon)\geq0$. Suppose that  $u$ is a solution of (\ref{VIGL}).
By the weak formulation
of (\ref{VIGL}), we get
$$\int_{\Omega_1}\left(|\nabla u|^2-\frac1{\varepsilon^2}(1-u^2)u^2\right)\md x
+\int_{\Omega_2}\left(\frac1m|\nabla u|^2+\frac{a}{\varepsilon^2}\,u^2\right)\md x=0.$$
Using the
identity $-(1-u^2)u^2=\frac12(1-u^2)^2-\frac12(1-u^4)$, we obtain from the preceding equation,
$$0=\mathcal G_0(u)-\frac{|\Omega_1|}{2\varepsilon^2}+\frac1{2\varepsilon^2}\int_{\Omega_1}u^4\,\md x.$$
Noticing that
$$\mathcal G_0(u)\geq \lambda_1(a,m,\varepsilon)\int_\Omega u^2\,\md
x+\frac{|\Omega_1|}{2\varepsilon^2}+\frac1{2\varepsilon^2}
\int_{\Omega_2}u^4\,\md x,$$
we get finally that
$$\int_{\Omega_1}u^4\,\md x=0.$$
Combined with Lemma~~\ref{VI-lemSec3}, we obtain that $u\equiv0$ in $\Omega$.\\
Suppose now that $\lambda_1(a,m,\varepsilon)<0$. Let $\varphi$ be a
normalized (in $L^2(\Omega;\mathbb R)$) eigenfunction corresponding
to $\lambda_1(a,m,\varepsilon)$. Then, for $\delta>0$, one has,
$$\mathcal
G_0(\delta\varphi)=\delta^2\left(\lambda_1(a,m,\varepsilon)+\delta^2
\frac1{2\varepsilon^2}\int_{{\Omega_1}}|\varphi|^4\,\md x\right)+\frac{|{\Omega_1}|}{2\varepsilon^2}.$$ By
choosing $\delta$ small enough, one gets,
$$\mathcal G_0(\delta\varphi)<\frac{|{\Omega_1}|}{2\varepsilon^2},$$
and consequently,  by the definition of $C_0(\varepsilon)$,
\begin{equation}\label{VI-Beg}
C_0(\varepsilon)< \frac{|{\Omega_1}|}{2\varepsilon^2}. \end{equation}
Since, $\mathcal G_0(0)=\frac{|{\Omega_1}|}{2\varepsilon^2}$, we get
that  $u\equiv0$ is not a minimizer
of $\mathcal G_0$.\hfill$\Box$\\

In the next proposition, we determine the minimizers of the functional
$\mathcal G_0$.

\begin{prop}\label{VI-LuPaProp2.1'}
Suppose that $\partial{\Omega_1},\partial\Omega$ are of class $C^3$,
and that $\lambda_1(a,m,\varepsilon)<0$. Then Equation (\ref{VIGL})
admits a non-negative non-trivial
solution. This solution is unique and satisfies,
\begin{enumerate}
\item $0<u_\varepsilon(x)<1$ on $\overline{\Omega}$;
\item The only minimizers of (\ref{VI-EnGL1}) are
$u_\varepsilon$ and $-u_\varepsilon$.
\end{enumerate}
\end{prop}
\paragraph{\bf Proof.}\ \\
{\it Step 1. Existence of a non-negative non-trivial solution.}\\
Let $u$ be a minimizer of (\ref{VI-EnGL1}). By
Proposition~\ref{VIVI-LuPaP2.1}, $u\not\equiv0$.  Let $v=|u|$. Then
$v\geq0$  is also a non-trivial minimizer of (\ref{VI-EnGL1}), and
hence a weak solution of
(\ref{VIGL}).\\
{\it Step~2. A non-negative non-trivial solution of (\ref{VIGL})  is positive.}\\
Let $v\geq0$ be a non-trivial solution of (\ref{VIGL}). By the
standard interior 
regularity theory, $v\in C^\infty(\Omega_1\cup\Omega_2)$. By
Theorem~\ref{VI-regKa} and the Sobolev imbedding theorem, we get,
thanks to the smoothness of the boundary,
$$v_{|_{{\Omega_1}}}\in C^{1,\alpha}(\overline{{\Omega_1}}),\quad
v_{|_{{\Omega_2}}}\in
C^{1,\alpha}(\overline{\Omega_2}),\quad \forall\alpha\in[0,1[.$$  We
claim that~:
\begin{equation}\label{VI-Claim1}
v>0,\quad \text{ in }\overline\Omega.
\end{equation}
Suppose by contradiction that that there exists
$x_0\in\overline\Omega$ such that $v(x_0)=0$. Notice that, we
have,
\begin{equation}\label{VI-MaxP1}
-\Delta v+c(x)v\geq 0\quad \text{ in }{\Omega_1},\quad -\Delta v+\frac{am}{\varepsilon^2}\,v=0\quad\text{ in }\Omega_2, \end{equation}
where $c(x)=(1/\varepsilon^2)v(x)^2\geq0$. If $x_0\in\Omega_1\cup\Omega_2$,
we get by the Strong Maximum Principle (Theorem~\ref{VI-MP-GiTr}-(2)),
$$\text{ either }v\equiv0\quad \text{in }
{\Omega_1},\quad\text{ or }v\equiv0\quad\text{in }\Omega_2.$$ Coming back to
Lemma~\ref{VI-lemSec3}, this yields a
contradiction.\\
If, otherwise, $x_0\in\partial{\Omega_1}$, then since $v$ satisfies 
(\ref{VI-MaxP1}), we get by the Hopf
Lemma (Theorem~\ref{VI-MP-GiTr}-(1)),
\begin{equation}\label{VI-Hopf}
\mathcal T_{\partial{\Omega_1}}^{\rm int}(\nu_1\cdot\nabla v)<0,\quad \mathcal
T_{\partial{\Omega_1}}^{\rm ext}(\nu_1\cdot\nabla v)>0 \quad\text{at }x_0,
\end{equation}
which contradicts the boundary condition in (\ref{VIGL}).
Therefore, the only possible choice is that $x_0\in\partial\Omega$,
but in this case
we get by the Hopf Lemma  a contradiction to the Neumann boundary
condition in (\ref{VIGL}).  We have thus proved Claim~(\ref{VI-Claim1}).\\
We claim now that $v<1$ in $\overline\Omega$. Let $x_0\in\overline\Omega$ be a maximum point of $v$,
$$v(x_0)=\max_{x\in \overline\Omega} v(x).$$
Suppose by contradiction that $v(x_0)\geq1$. Let $w=1-u^2$.
Since
$$\Delta (u^2)=2u\Delta u+2|\nabla u|^2,\quad
\nabla (u^2)=2u\nabla u,$$
the function $w$ satisfies~:
$$-\Delta w+c(x)\,w\geq 0\quad\text{in }\Omega_1,\quad
-\Delta w+\frac{2}{\varepsilon^2}am\,w\geq0\quad\text{ in }\Omega_2,$$
together with the boundary conditions:
$$\mathcal T_{\partial\Omega_1}^{\rm int}\left(\frac{\partial w}{\partial\nu_1}\right)=\frac1m
\mathcal T_{\partial\Omega_1}^{\rm ext}\left(\frac{\partial w}{\partial\nu_1}\right),\quad
\mathcal T_{\partial\Omega}^{\rm int}\left(\frac{\partial w}{\partial\nu}\right)=0,$$
and
$$c(x)=\frac{2}{\varepsilon^2}u(x)^2\geq0,\quad w(x_0)=\min_{x\in\overline\Omega}w(x)\leq0.$$
Then, as for the proof of Claim (\ref{VI-Claim1}), we get a contradiction by Theorem~\ref{VI-MP-GiTr}.\\
{\it Step~3. The positive solution is unique.}\\
We now claim that the positive solution $u$ obtained in Steps~1~and~2 above
is unique.
It is sufficient to prove the following claim~:
\begin{equation}\label{VI-u1u2}
\text{If } u_1\text{ and } u_2  \text{ are positive solutions of }(\ref{VIGL}),
\text{ then }u_1\geq u_2.
\end{equation}
To prove Claim (\ref{VI-u1u2}), we shall follow the argument of
Lu-Pan~\cite{LuPa96}. For $\lambda\geq1$, we denote by
$u_\lambda=\lambda u_1$. Since $u_1,u_2>0$ in
$\overline{\Omega}$ and $\Omega$ is bounded, then for $\lambda$ large enough, we
have, $u_\lambda>u_2$. Let us define the following number,
$$\bar\lambda=\inf\left\{\,\lambda\geq1~:~u_\lambda\geq
u_2\quad\text{in }~\overline{\Omega}\,\right\}.$$ Then it is
sufficient to prove that $\bar\lambda=1$. Suppose by contradiction
that $\bar\lambda>1$. Then $\bar u:=u_{\bar\lambda}$ satisfies,
\begin{equation}\label{VI-prop-u}
\bar u\geq u_2,\quad \inf_{x\in\overline{\Omega}}(\bar
u-u_2)=0,
\end{equation}
and $\bar u$ is a super-solution of (\ref{VIGL}), i.e.
\begin{equation}\label{VI-SupSolGL}
\left\{\begin{array}{l} -\Delta\bar u\geq\displaystyle\frac1{\varepsilon^2}(1-\bar
u^2)\bar u,\quad\text{in }{\Omega_1},\\
\\
-\Delta\bar u+\displaystyle\frac1{\varepsilon^2}am\, \bar u\geq0,\quad\text{in }\Omega_2,\\
\\
\mathcal T_{\partial{\Omega_1}}^{\rm int}(\nu_1\cdot\nabla \bar u)=\frac1m\mathcal
T_{\partial{\Omega_1}}^{\rm ext}(\nu_1\cdot\nabla\bar u),\quad \text{on }\partial{\Omega_1},\\
\\
\mathcal T_{\partial\Omega}^{\rm int}(\nu\cdot\nabla
\bar u)=0,\quad\text{on }\partial\Omega.
\end{array}\right.
\end{equation}
Let $x_0\in\overline{\Omega}$ be such that $(\bar
u-u_2)(x_0)=0$. Let $c_1(x)=\left[(\bar u^2+\bar
uu_2+u_2^2)(x)\right]/\varepsilon^2$, then $c_1(x)> 0$ and we have,
$$-\Delta (\bar u-u_2)+c_1(x)(\bar u-u_2)\geq 0\quad\text{in }{\Omega_1},\quad
-\Delta(\bar u-u_2)+am(\bar u-u_2)\geq0\quad\text{in }\Omega_2.$$ By
the Strong Maximum Principle, we get that
$x_0\in\partial{\Omega_1}\cup\partial\Omega$. As in Step~2
before, we get using Hopf's Lemma and the boundary conditions
satisfied by $u_2$ and $\bar u$ that this case is
impossible.\hfill$\Box$\\

\paragraph{\bf Proof of Theorem~\ref{VI-LuPaTh1}.}\ \\
The Assertions (1) and (2) are consequences of 
Propositions~\ref{VIVI-LuPaP2.1}~and~\ref{VI-LuPaProp2.1'}.\\
{\it Proof of Assertion (3).}\\ 
After a  Coulomb gauge transformation
(cf.~\cite{BR}) we can look for minimizers of (\ref{VI-EGL}) in the
space $H^1(\Omega;\mathbb C)\times H^1_{\rm div}(\Omega;\mathbb R^2)$, where
\begin{equation}\label{VI-GT}
H_{\rm div}^1(\Omega;\mathbb R^2)=\{A\in H^1(\Omega;\mathbb R^2)~:~{\rm div}\,A=0\quad\text{in }\Omega,\quad \nu\cdot
A=0\quad\text{on }\partial\Omega\}.
\end{equation} 
The existence of minimizers of
(\ref{VI-EGL}) is then standard starting from a minimizing sequence in
the space  $H^1(\Omega;\mathbb C)\times H^1_{\rm div}(\Omega;\mathbb R^2)$
(cf.~\cite{Gi, LuPa96}).\\
Let $(\psi,A)$ be a minimizer of (\ref{VI-EGL}). To prove assertion (3) of Theorem~\ref{VI-LuPaTh1}, it is sufficient
to prove that $|\psi|$ is a minimizer of (\ref{VI-EnGL1}). Notice
that, by Kato's inequality (cf.~\cite[Proposition~6.6.1]{JT}), we
have,
$$\int_{\Omega}|(\nabla-iA)\psi|^2\,\md
x\geq\int_{\Omega}\left|\nabla|\psi|\,\right|^2\md x,$$ 
which implies (recall that $H=0$),
\begin{equation}\label{VI-LB-En}
\mathcal G(\psi,A)\geq\mathcal G_0(|\psi|)+\int_{\Omega_1}|{\rm
curl}\,A|^2\,\md x+\frac1\mu\int_{\Omega_2}|{\rm curl}\,A|^2\,\md x.
\end{equation}
On the other hand,  for a minimizer $u_\varepsilon$ of
(\ref{VI-EnGL1}), we have,
$$\mathcal G(\psi,A)=
\inf_{(\phi,B)\in H^1(\Omega;\mathbb C)\times H^1(\Omega;\mathbb
R^2)} \mathcal G(\phi,B)\leq\mathcal G(u_\varepsilon,0)=\mathcal
G_0(u_\varepsilon)=\inf_{v\in H^1(\Omega)}\mathcal G_0(v).$$
Combined with (\ref{VI-LB-En}), this permits us to deduce that
\begin{equation}\label{VI-G=G0}
\mathcal G_0(|\psi|)=\mathcal G_0(u_\varepsilon). \end{equation}
Hence, $|\psi|$ is a minimizer of (\ref{VI-EnGL1}) and
consequently,
by Proposition~\ref{VI-LuPaProp2.1'}, $|\psi|\equiv u_\varepsilon$.\\
{\it Proof of Assertion (4).}\\
If $\lambda_1(a,m,\varepsilon)\geq0$, then by Proposition~\ref{VIVI-LuPaP2.1}, $u_\varepsilon\equiv0$ and
we have nothing to prove. So suppose that 
$\lambda_1(a,m,\varepsilon)<0$
(i.e. $u_\varepsilon>0$). Since $\Omega$ is bounded and  simply connected, and 
$|\psi|=u_\varepsilon$, then it is  a general  result in \cite{BBM}
(see also \cite{BZ}) that there exists a `lift'  $\chi\in
H^1(\Omega;\mathbb R)$
(unique up to $2k\pi$, $k\in\mathbb Z$) such that,
$$\psi=u_\varepsilon e^{i\chi}.$$
It is sufficient then to prove that $A=\nabla\chi$.
Notice that we have (since ${\rm curl}\,A=0$),
\begin{eqnarray}\label{VI-REV1}
\mathcal G(\psi,A)&=&\int_{\Omega_1}\left(|\nabla
u_\varepsilon|^2+u_\varepsilon^2|\nabla\chi-A|^2+\frac1{2\varepsilon^2}
\left(1-|u_\varepsilon|^2\right)^2\right)\md
x\\
&&+\int_{\Omega_2}\left(\frac1m\left(|\nabla
u_\varepsilon|^2+u_\varepsilon^2|\nabla\chi-A|^2\right)+\frac{a}{\varepsilon^2}|u_\varepsilon|^2\right)\md
x.\nonumber
\end{eqnarray}
Therefore, when combined with (\ref{VI-G=G0}), (\ref{VI-REV1}) yields,
$$\int_{\Omega_1}|\nabla\chi-A|^2u_\varepsilon^2\,\md
x+\frac1m\int_{\Omega_2}|\nabla\chi-A|^2u_\varepsilon^2\,\md x=0.$$
By Proposition~\ref{VI-LuPaProp2.1'}, $u_\varepsilon>0$ and
consequently $A=\nabla\chi$. Since $A\in H^1(\Omega;\mathbb R^2)$, it
follows that $\chi\in H^2(\Omega)$ thus achieving the proof of
Theorem~\ref{VI-LuPaTh1}.
\hfill$\Box$\\

\section{Existence and uniqueness in $\mathbb R^2$}\label{VISec4}
In this section, we prove Theorem~\ref{VI-sol-modeq}. That is, in the
class of functions
(\ref{VI-C}),
Equation (\ref{VI-limeq}) admits a unique solution, which is given by (\ref{VI-Sol}).\\
Let us explain how we have obtained the expression of the solution
(\ref{VI-Sol}). As in \cite{LuPa96}, we look for a solution of
(\ref{VI-limeq}) in the form~:
$$(x_1,x_2)\mapsto U(x_2).$$
Then $U$ is a solution of the following  ODE~:
\begin{equation}\label{VI-ODE}
\left\{\begin{array}{l} -U''=(1-U^2)U,\quad x_2>0,\\
-U''+am\,U=0,\quad x_2\leq0,\\
U'(0_+)=\frac1m\,U'(0_-),\quad U(0_+)=U(0_-).
\end{array}\right.
\end{equation}
Assuming that $U$ is bounded, the second equation in (\ref{VI-ODE})
gives that,
$$U(x_2)=A\,\exp\left(\sqrt{am}\,x_2\right),\quad x_2<0,\quad A>0.$$
We obtain now from equation (\ref{VI-ODE}),
\begin{equation}\label{VI-LuPa-limeq}
\left\{
\begin{array}{l}
-U''=(1-U^2)U,\quad x_2>0,\\
U'(0_+)=\sqrt{2}\ell\,U(0),
\end{array}\right.
\end{equation}
where
\begin{equation}\label{VI-LuPa-ell}
\ell=\sqrt{\frac{a}{2m}}.
\end{equation}
The  positive solution of
(\ref{VI-LuPa-limeq}) is unique and  is given by (see \cite[Section~5]{LuPa96})~:
$$U(x_2)=\frac{\beta\exp(\sqrt{2}x_2)-1}{\beta\exp(\sqrt{2}x_2)+1},$$
with $\beta=\frac{1+\sqrt{1+\ell^2}}{\ell}$.\\
Using the boundary
condition $U(0_+)=U(0_-)$, we get,
$$A=\frac{\beta-1}{\beta+1}=\frac{\sqrt{2m}+\sqrt{a+2m}-\sqrt{a}}{\sqrt{2m}+\sqrt{a+2m}+\sqrt{a}}.$$

The uniqueness and the symmetry of positive solutions to semilinear elliptic equations in a
half-space $\mathbb R_+^n$ with either Dirichlet or Robin (de\,Gennes) boundary condition  on $\mathbb R^{n-1}\times\{0\}$
have been studied extensively (cf.~\cite{BCN, D, GNN, G, LuPa96}). To
prove Theorem~\ref{VI-sol-modeq}, we shall use methods inspired from these
papers and mainly from \cite{LuPa96}. The main technical difficulty
is due to the singularity of the solutions on the boundary $\mathbb R\times\{0\}$.
%If one want to use the argument of Lu-Pan~\cite{LuPa96}, then given a
%non-trivial bounded solution of (\ref{VI-limeq}), we have to establish
%the following properties~:
%\begin{enumerate}
%\item $0<u(x)<1$ for all $x\in\mathbb R^2$;
%\item There exists a constant $\lambda\in]0,1]$ such that~:
%$$\lambda u(x)\leq  U(x),\quad \lambda U(x)\leq u(x),\quad \forall x\in \mathbb R^2.$$
%\item The optimal constant $\lambda_*$ for which the property (2) holds is actually $\lambda_*=1$.
%\end{enumerate}
%Note that property (3) above is sufficient to deduce the conclusion of Theorem~\ref{VI-sol-modeq}. We
%are able to establish the property (1) for equation (\ref{VI-limeq}), see Lemma~\ref{VI-R2-0<u<1},
%but we were not able to prove
%(2) for $x\in\mathbb R\times\mathbb R_-$ and hence to deduce property (3). However, for the purpose of the
%equation (\ref{VI-limeq}), we have to prove first that $u\equiv U$ in
%$\mathbb R\times\mathbb R_+$ (by proving similar
%properties to (2)-(3)) and then we prove seperately that $u\equiv U$ in $\mathbb R\times\mathbb R_-$.\\

A first step  is the analysis of the following linear equation~:
\begin{equation}\label{VI-lineq}
-\Delta u+\alpha u=0,\quad\text{ in }\mathbb R^2.
\end{equation}
The next lemma is well known. We include a
proof for the reader's convenience, which illustrates in a simple case
the arguments that will be used later.
\begin{lem}\label{VI-Lemlineq}
Suppose that $\alpha>0$. If $u\geq 0$ is a bounded strong solution of (\ref{VI-lineq}), then $u\equiv0$.
\end{lem}
\paragraph{\bf Proof.}
We define
$$\widetilde M=\sup_{x\in\mathbb R^2}u(x).$$
We shall prove that $\widetilde M=0$. Suppose by contradiction that
$\widetilde M>0$. Notice that the Strong Maximum Principle yields~:
\begin{equation}\label{VI-MPR2}
u(x)<\widetilde M\quad \text{ in } \mathbb R^2.
\end{equation}
Let $\{x^n\}_{n=1}^\infty\subset \mathbb R^2$ be
a sequence such that
$$\lim_{n\to+\infty}u(x^n)=\widetilde M.$$
By (\ref{VI-MPR2}), we get that the sequence  $\{x^n\}$ is unbounded
and hence we may extract from it a subsequence, still denoted by $x^n$, such that
$|x^n|\to+\infty$. Let us define the function~:
$$u_n(x)=u(x+x^n),\quad\forall x\in\mathbb R^2.$$
Then $u_n$ is a solution of (\ref{VI-lineq}) and
$\|u_n\|_{L^\infty(\mathbb R^2)}\leq \widetilde M$. We claim that
there exist a subsequence of $u_n$ (still denoted by $u_n$) and a function $\widetilde u\in C^2(\mathbb R^2)$ such that~:
\begin{equation}\label{VI-compact-argument}
u_n\to \widetilde u\quad \text{in }C_{\rm loc}^2(\mathbb R^2).
\end{equation}
Here we mean by convergence in $C_{\rm loc}^2$, that for any compact
subset $K\subset \mathbb R^2$, $(u_n)_{|_{K}}$ converges to
$\widetilde u_{|_{K}}$ in $C^2(K)$. To prove (\ref{VI-compact-argument}), let $R>0$ and
$D_R$ the open disc centered at $0$ and of radius $R$. 
By the elliptic estimates and
the Sobolev Imbedding Theorem, we get a constant $C_R>0$ such that
$$\|u_n\|_{H^4(D_R)}\leq C_R,\quad\forall~n\in\mathbb N.$$
Since the space 
$H^4_{\rm loc}(\mathbb R^2)$ is compactly imbedded in $C^2_{\rm
  loc}(\mathbb R^2)$ 
(cf. \cite[Theorem~7.26]{GiTr}), we get that the sequence $u_n$ is
precompact in $C_{\rm loc}^2(\mathbb R^2)$. This proves (\ref{VI-compact-argument}).\\  
Notice that $\widetilde u$ is also a solution of (\ref{VI-lineq}), $0\leq\widetilde u\leq \widetilde M$
and $\widetilde u(0)=\widetilde M$. Therefore, by the Strong Maximum Principle, we get that $\widetilde u\equiv\widetilde M$,
which is not a solution of (\ref{VI-lineq}) unless $\widetilde M=0$.\hfill$\Box$\\

\begin{lem}\label{VI-R2-0<u<1}
Let $u$ be a bounded strong solution of (\ref{VI-limeq}), $u\geq 0$ and $u\not\equiv0$. Then $0<u(x)<1$ for all
$x\in\mathbb R^2$.
\end{lem}
\paragraph{\bf Proof.}\ \\
{\it Step 1. $u>0$ in $\mathbb R^2$.}\\
This follows from Theorem~\ref{VI-MP-GiTr}, exactly as in the
bounded case (Proof of Proposition~\ref{VI-LuPaProp2.1'}, Step~2).\\
{\it Step 2. $u\leq 1$ in $\mathbb R^2$.}\\
We denote by
$$M_-=\sup_{x\in\mathbb R\times\mathbb R_-}u(x),\quad
M_+=\sup_{x\in\mathbb R\times\mathbb R_+}u(x),\quad M=\max(M_-,M_+).$$
It is sufficient to show that $M\leq 1$. The proof is twofold, whether $M=M_-$ or $M=M_+$.\\
{\it Case 2.1. $M=M_-$ (i.e. $M_+\leq M_-$).}\\
Suppose by contradiction that $M_->1$.
Let $x^n=(x_1^n,x_2^n)$ be a sequence in $\mathbb R\times\mathbb R_-$ such that
$$\lim_{n\to+\infty}u(x^n)=M_-.$$
We make the following claim~:
\begin{equation}\label{VI-Claim1.a}
\exists\, \delta>0,\quad \limsup_{n\to+\infty}x_2^n\leq-2\delta.
\end{equation}
We define the following function~:
$$u_n(x_1,x_2)=u(x_1+x_1^n,x_2+x_2^n),\quad\forall x\in\mathbb R^2.$$
By the claim (\ref{VI-Claim1.a}), we get
$$-\Delta u_n+am\, u_n=0\quad\text{in }D_{3\delta/2},$$
where, for $r>0$, $D_{r}\subset\mathbb R^2$ denotes the open  disc of
center $0$ and radius $r$. Using the argument of the proof of
(\ref{VI-compact-argument}), we get a function $\widetilde u\in
C^2(D_{3\delta/2})$ and a subsequence of $u_n$ that converges to
$\widetilde u$ in $C_{\rm loc}^2(D_{3\delta/2})$. In particular, the function $\widetilde u$ satisfies~:
$$-\Delta \widetilde u+am\,\widetilde u=0\text{ in }D_\delta,\quad
0\leq \widetilde u\leq M_-,\quad \widetilde u(0)=M_-.$$
By the Strong Maximum Principle, we obtain that $\widetilde u\equiv M_-$. Coming back to the equation satisfied
by $\widetilde u$ we get that $M_-=0$ which is the desired
contradiction. Therefore, the only missing  point  is the proof
of Claim (\ref{VI-Claim1.a}).\\
Suppose by contradiction that there exists a subsequence of $x^n_2$ (still denoted by $x_2^n$) such that
$$\lim_{n\to+\infty}x_2^n=0.$$
We define the function $v_n(x_1,x_2)=v(x_1+x_1^n,x_2)$ $(x=(x_1,x_2)\in\mathbb R^2)$. It is clear that $v_n$
is a solution of (\ref{VI-limeq}). We can extract a subsequence
of $v_n$ that converges to a function $v$ in $C^2_{\rm loc}(\overline{\mathbb
R\times\mathbb R_\pm})$.
Notice that:
\begin{itemize}
\item $v$ is a solution of (\ref{VI-limeq});
\item $0\leq v\leq M_-$ in $\mathbb R^2$;
\item $v(0)=M_-$;
\item Writing $w(x)=1-v(x)$, $c(x)=(1+v(x))v(x)$, we get
\begin{equation}\label{VI-mp100}
w(0)\leq w(x),\quad c(x)\geq 0,\quad -\Delta w+c(x)w=0\text{ in }\mathbb R\times\mathbb R_+.
\end{equation}
\end{itemize}
Therefore, we get the following two inequalities~:
$$\left(\frac{\partial w}{\partial x_2}\right)(0,0_-)\geq 0,\quad
\left(\frac{\partial w}{\partial x_2}\right)(0,0_+)<0.$$
The first inequality is an immediate consequence of the fact that $v$ attains a maximum at $0$, and the second
is nothing but the Hopf Lemma (Theorem~\ref{VI-MP-GiTr}-(1))  applied to the function $w$
(cf.~(\ref{VI-mp100})). 
Coming back to the boundary condition satisfied by $v$, we arrive at the desired contradiction.\\
{\it Case 2.2. $M=M_+$ (i.e. $M_-\leq M_+$).}
The proof is just as in Case~2.1 (details are given in~\cite[Lemma~5.2]{LuPa96}).\\
{\it Step 3. $u(x)<1$ in $\mathbb R^2$.}\\
Suppose by contradiction that there exists $x_0\in\mathbb R^2$ such that $u(x_0)=1$. The Strong Maximum Principle
yields $x_0\not\in\mathbb R\times\mathbb R_-$. The Hopf Lemma and the boundary condition satisfied by $u$ yield also that
$x_0\not\in\mathbb R\times\{0\}$. So 
$x_0\in\mathbb R\times\mathbb  R_+$. Putting $c(x)=(1+u)u(x)$ and $w(x)=1-u(x)$, we get
$$-\Delta w+c(x)w\geq 0,\quad\text{ in }\mathbb R\times\mathbb R_+,$$
with $c(x)\geq 0$. The Strong Maximum Principle now gives $w\equiv0$ in $\mathbb R\times\mathbb R_+$ (i.e. $u\equiv0$).
Coming back to the equation satisfied by $u$ in $\mathbb R\times\mathbb R_-$ and the boundary condition, we obtain
$$-\Delta u+am u=0\text{ in }\mathbb R\times\mathbb R_-,\quad\frac{\partial u}{\partial x_2}(\cdot,0_-)=0
\text{ in }\mathbb R.$$
We define now the  function $\widetilde u$ in $\mathbb R^2$ by~:
$$\widetilde u(x_1,x_2)=u(x_1,-x_2)\text{  if }x_2>0,\quad
\widetilde u(x_1,x_2)=u(x_1,x_2)\text{ if }x_2<0.$$
We then get that $\widetilde u$ is a weak solution (by elliptic regularity theory it becomes a strong solution)
of Equation (\ref{VI-lineq}) with $\alpha=am$. By Lemma~\ref{VI-Lemlineq}, we get that $\widetilde u\equiv0$. Therefore,
we obtain finally~:
$$u\equiv 0\quad\text{ in }\mathbb R\times\mathbb R_-,\quad u\equiv1\text{ in }
\mathbb R\times\mathbb R_+,$$
which is the desired contradiction.\hfill$\Box$\\

\begin{lem}\label{VI-unifLB}
Given $a,m>0$, there exist constants $C_-,C_+\in]0,1[$ such that, if  $u>0$ is a bounded strong solution of (\ref{VI-limeq}), then,
\begin{equation}\label{VI-unifLBeq}
\sup_{x\in\mathbb R\times\mathbb R_-}u(x)<1-C_-,\quad\inf_{x\in\mathbb R\times\mathbb R_+}u(x)>C_+.
\end{equation}
\end{lem}
\paragraph{\bf Proof.}\ \\
{\it Step~1. Existence of $C_-$.}\\
Suppose by contradiction that there exist sequences $u_n$ and $x^n=(x_1^n,x_2^n)\in\mathbb R\times\mathbb R_-$ such that
$u_n\geq0$ is a bounded strong solution of (\ref{VI-limeq}) and
$$\lim_{n\to+\infty} u_n(x^n)=1.$$
We define the function $\bar u_n(x_1,x_2)=u_n(x_1+x_1^n,x_2)$. Then $\bar u_n$ is a solution of (\ref{VI-limeq})
and $\displaystyle\lim_{n\to+\infty}\bar u_n(0,x_2^n)=1$.\\
We claim that $x_2^n$ is unbounded. If not, then we may extract a subsequence (still denoted by $x^n_2$) such that
$\displaystyle\lim_{n\to+\infty} x_2^n=b$ for some $b\leq0$. As in the
proof of Lemma~\ref{VI-Lemlineq}, we show that there exists  a function
$\bar u$ such that a subsequence of
$\bar u_n$ converges to $\bar u$ in $C^2_{\rm loc}(\overline{\mathbb
  R\times\mathbb R_\pm})$. Notice that $\bar u$ is a solution of (\ref{VI-limeq})
and $\bar u(0,b)=1$. Putting
$$\bar w=1-\bar u,\quad c(x)=(1+\bar u(x))\bar u(x),$$
then
$$-\Delta \bar w+am\,\bar w\geq0\quad\text{in }\mathbb R\times\mathbb R_-,\quad
-\Delta \bar w+c(x)\bar w=0\quad\text{in }\mathbb R\times\mathbb R_+.$$
If $b<0$, we get a contradiction by the Strong Maximum
Principle. So $b=0$.
By Hopf Lemma, we get~:
$$\frac{\partial\bar w}{\partial x_2}(0,b_-)<0,\quad \frac{\partial\bar w}{\partial x_2}(0,b_+)>0.$$
Coming back to the boundary condition satisfied by $\bar u$ (cf. (\ref{VI-limeq})), we get the desired contradiction.\\
Therefore, having proved that $\displaystyle \lim_{n\to+\infty}x_2^n=-\infty$, we define the following function~:
$$w_n(x_1,x_2)=\bar u_n(x_1,x_2+x_2^n),\quad
\forall~(x_1,x_2)\in\mathbb R^2.$$
Notice that, there exists $n_0>0$ large enough so that
$$\forall~n\geq n_0,\quad-\Delta w_n+am\,w_n=0\quad \text{ in }D_1,$$
where $D_1$ is the unit open disc.\\
Since $\|w_n\|_{L^\infty(\mathbb R^2)}\leq1$, we get by the elliptic
estimates and the Sobolev Imbedding Theorem a subsequence of
$w_n$ that converges to a function $w$ in $C^2(D_{1/2})$. Moreover,
$w$ satisfies,
\begin{equation}\label{VI**}
-\Delta w+am\,w=0\text{ in }D_{1/2},\quad
0\leq w\leq 1,\end{equation}
and $w(0)=1$. By the Strong Maximum principle, we get that $w\equiv1$
in $D_{1/2}$, which is not a solution  of (\ref{VI**}) and so we get a contradiction.
Therefore, we have proved the existence of $C_-$.\\
{\it Step~2. Existence of $C_+$.}\\
The argument is also by contradiction, but we shall not give the details
refering the reader to \cite[Lemma~5.3]{LuPa96}.\hfill$\Box$\\

\begin{lem}\label{VI-limx2}
Let $u>0$ be a bounded strong solution of (\ref{VI-limeq}).
Then
the following limits hold~:
\begin{equation}\label{VI-limx2-eq}
\lim_{x_2\to-\infty}\left(\sup_{x_1\in\mathbb R}u(x_1,x_2)\right)=0,\quad
\lim_{x_2\to+\infty}\left(\sup_{x_1\in\mathbb R}(1-u(x_1,x_2))\right)=0.
\end{equation}
\end{lem}
\paragraph{\bf Proof.}\ \\
We give the proof of the limit as $x_2\to-\infty$.
Suppose by contradiction that there exists $\epsilon>0$ and a sequence $(x_1^n,x_2^n)\in\mathbb R\times\mathbb R_-$
such that~:
$$\lim_{n\to+\infty}x_2^n=-\infty,\quad \text{and }\epsilon<u(x_1^n,x_2^n).$$
Let us consider the sequence of functions 
$u_n(x_1,x_2)=u(x_1+x_1^n,x_2+x_2^n)$. Then, given $R>0$, there exists
$n_0\in\mathbb N$ such that~:
$$\forall~n\geq n_0,\quad-\Delta u_n+am\,u_n=0\quad \text{in } D_R.$$
Again, since $\|u_n\|_{L^\infty(D_R)}\leq 1$, we get by the elliptic
estimates and the Sobolev
Imbedding Theorem a subsequence of $(u_n)$ that converges to a function
$\bar u$ in $C_{\rm loc}^2(\mathbb R^2)$. The function $\bar u$ is a solution of Equation (\ref{VI-lineq}) (with $\alpha=am$)
and $\bar u(x_1,0)\geq\epsilon$. By Lemma~\ref{VI-Lemlineq}, we get $\bar u\equiv0$, which is the desired contradiction.\\
The proof when $x_2\to+\infty$ is exactly as that given in \cite[(5.9)]{LuPa96}.
%{\it Proof of assertion (2).}\\
%Let us consider a sequence $x^n=(x_1^n,x_2^n)\in\mathbb R\times[-\delta,\delta]$ such that
%$$\lim_{n\to+\infty} |x_1^n|=+\infty,\quad\lim_{n\to+\infty}x_2^n=b,$$
%for some $b\in[-\delta,\delta]$. It is sufficient to prove that~:
%$$0<\displaystyle\lim_{n\to+\infty}u(x^n)<1.$$
%Let us prove that $\displaystyle\lim_{n\to+\infty}u(x^n)>0$ (the proof of $\displaystyle\lim_{n\to+\infty}u(x^n)<1$ %will be the same). Suppose by contradiction that $\displaystyle\lim_{n\to+\infty}u(x^n)=0$.
%We can extract a subsequence of $x\mapsto u(x_1+x^n_1,x_2)$ that converges to some function $\bar u$ in $C_{\rm %loc}^2(\mathbb R\times[-\delta,\delta])$.
%Notice that $\bar u$ is a solution of (\ref{VI-limeq}) in $\mathbb R\times]-\delta,\delta[$ and that
%$\bar u(0,b)=0$. This is impossible, thanks to Hopf Lemma and the boundary condition satisfied by $\bar u$, unless
%$\bar u\equiv0$.
\hfill$\Box$\\

The next lemma remains an essential step towards the proof of Theorem~\ref{VI-sol-modeq}.

\begin{lem}\label{VI-lambda*=1}
Let $u_1,u_2>0$ be two bounded strong solutions of (\ref{VI-limeq}).
Suppose moreover that there exists $\lambda\in]0,1[$ such that we
have in $ \mathbb R\times\mathbb R_+$~:
$$(H_\lambda)~~\left\{
\begin{array}{l}
u_2(x)\geq \lambda u_1(x),\\
u_2(x_1,x_2)+\frac1mu_2(x_1,-x_2)\geq\lambda\left(u_1(x_1,x_2)+\frac1mu_1(x_1,-x_2)\right).
\end{array}\right.$$
Then the following two assertions hold
\begin{enumerate}
\item
$u_2(x)>\lambda u_1(x)$ in $\overline{\mathbb R\times\mathbb R_+}$;
\item $u_2(x)\geq u_1(x)$ in $\mathbb R\times\mathbb R_+$.
\end{enumerate}
\end{lem}
\paragraph{\bf Proof.}\ \\
Let us establish Assertion (1). We denote by~:
\begin{equation}\label{VI-w(x)}
w_\lambda(x)=u_2(x)-\lambda u_1(x),\quad\forall x\in\mathbb R\times\mathbb R.
\end{equation}
Notice that, by hypothesis, $w_\lambda\geq0$ and it satisfies~:
$$-\Delta w_\lambda+c(x)w_\lambda\geq 0,\quad\text{ in }\mathbb R\times\mathbb R_+,$$
where $c(x)=\left(u_2^2+\lambda u_1u_2+\lambda^2u_1^2\right)(x)\geq0$. By the Strong Maximum Principle, we get
that $w_\lambda>0$ in $\mathbb R\times\mathbb R_+$. So it remains to prove that $w_\lambda>0$ on $\mathbb R\times\{0\}$.
We define the function $h_\lambda$ on $\mathbb R\times\mathbb R_+$ by~:
$$h_\lambda(x_1,x_2)=w_\lambda(x_1,x_2)+\frac1m w_\lambda(x_1,-x_2).$$
Notice that, thanks to the boundary conditions satisfied by $u_1$ and $u_2$,
\begin{equation}\label{VI-h(x)-ND}
\frac{\partial h_\lambda}{\partial x_2}(\cdot,0)=0\quad\text{ on }\mathbb R.
\end{equation}
It is easy to prove that $h_\lambda$ satisfies~:
$$-\Delta h_\lambda+(2\lambda +am)h_\lambda\geq0,\quad\text{ in }\mathbb R\times\mathbb R_+.
$$
So, if there exists $x_0\in\mathbb R\times\{0\}$ such that $w_\lambda(x_0)=0$, then Hopf's Lemma will give
$\frac{\partial h_\lambda}{\partial x_2}(x_0)>0$, which contradicts (\ref{VI-h(x)-ND}). Therefore, this proves that
$w_\lambda>0$ on $\mathbb R\times\{0\}$. This finishes the proof of Assertion (1) of the lemma.\\
Now we prove assertion (2). Let us define $\lambda_*$ by~:
$$\lambda_*=\inf\{\lambda\in]0,1];\quad (H_\lambda) \text{ holds in }\mathbb R\times\mathbb R_+\}.$$
It is sufficient to prove that $\lambda_*=1$. Suppose by contradiction that $\lambda_*<1$.
Let us write $w=w_{\lambda_*}$. Then $w$ satisfies~:
$$\inf_{x\in\mathbb R\times\mathbb R_+}w(x)=0,$$
and by Assertion (1), $w>0$ in $\overline{\mathbb R\times\mathbb R_+}$. Let $x^n\in\mathbb R\times\mathbb R_+$ be
a sequence such that $\displaystyle\lim_{n\to+\infty}w(x^n)=0$. Then one should have  $x^n$ unbounded.
So, we can suppose that $\displaystyle\lim_{n\to+\infty}|x^n|=+\infty$.\\
Now, $x^n_2$ should be bounded since, by Lemma~\ref{VI-limx2}, $\displaystyle\lim_{x_2\to+\infty}
w(x_1^n,x_2)=(1-\lambda)$ uniformly with respect to $x_1^n$. So we may suppose that $\displaystyle
\lim_{n\to+\infty}x_2^n=b$, for some $b\geq0$.\\
Thus, we have $\displaystyle\lim_{n\to+\infty}|x_1^n|=+\infty$. Let us define the function $u^n_1$ by~:
$$u^n_1(x_1,x_2)=u(x_1+x_1^n,x_2),\quad\forall (x_1,x_2)\in\mathbb R^2.$$
Then, there exists a subsequence of $u_1^n$ that converges to a function $\widetilde u_1$ in
$C^2_{\rm loc}(\overline{\mathbb R\times\mathbb R_\pm})$.
The function $\widetilde u_1$ is a strong, positive and bounded solution of (\ref{VI-limeq}) and it satisfies,
\begin{equation}\label{VI-w(0,b)=0}
(u_2-\lambda_*\widetilde u_1)(0,b)=0.
\end{equation}
Notice also that $\widetilde u_1,u_2$
satisfy the hypothesis $(H_{\lambda_*})$, hence, by assertion (1) of the lemma, we have
$u_2-\lambda_*\widetilde u_1>0$ in $\overline{\mathbb R\times\mathbb R_+}$, contradicting (\ref{VI-w(0,b)=0}).
Therefore, $\lambda_*=1$.
\hfill$\Box$\\

\paragraph{\bf Proof of Theorem~\ref{VI-sol-modeq}.}
Let $u\in\mathcal C$ (see (\ref{VI-C})) be a solution of (\ref{VI-limeq}). We shall prove that $u\equiv U$ by two steps~:
\begin{itemize}
\item First we establish that $u\equiv U$ in $\mathbb R\times\overline{\mathbb R_+}$.
\item Using the transmission conditions, we get sufficient information about $u$ on $\mathbb
  R\times\{0\}$ that permit us to establish 
that $u\equiv U$ in $\mathbb R\times\mathbb R_-$.
\end{itemize}
{\it Step 1. $u\equiv U$ in $\mathbb R\times\overline{\mathbb R_+}$.}\\
Let $u_1, u_2\geq0$ be two bounded solutions of (\ref{VI-limeq}). Notice that there exists $\lambda\in]0,1[$
such that $u_1,u_2$ satisfy the hypothesis $(H_\lambda)$. Actually, by Lemmas~\ref{VI-R2-0<u<1}~and~\ref{VI-unifLB},
it is sufficient to take~:
$$\lambda\in\left]0,\min\left\{1,\left(1+\frac1m(1-C_-)\right)^{-1}C_+\right\}\right],$$
where $C_-,C_+\in]0,1[$ are the constants of Lemma~\ref{VI-unifLB}.\\
Therefore, we obtain by Lemma~\ref{VI-lambda*=1} that $u_2\geq u_1$ in $\mathbb R\times\mathbb R_+$. Since the solutions
$u_1,u_2$ were arbitrarly chosen, this yields that $u\equiv U$ in $\mathbb R\times\mathbb R_+$.\\
{\it Step 2. $u\equiv U$ in $\mathbb R\times\mathbb R_-$.}\\
Let $u_1,u_2\geq0$ be again two solutions of (\ref{VI-limeq}). It is sufficient to prove that $u_2\geq u_1$
in $\mathbb R\times\mathbb R_-$. Notice that by Lemma~\ref{VI-unifLB},
we get for $\lambda\in]0,C_-]$,
$$(H'_\lambda)\quad(1-u_1)(x)\geq \lambda (1-u_2)(x)\quad \text{ in }\mathbb R\times\mathbb R_-.$$
Notice that  if $u_1,u_2$ satisfy the hypothesis $(H'_\lambda)$ for
some $\lambda\in]0,1[$, then 
$$1-u_1>\lambda(1-u_2),\quad \text{ in }
\mathbb R\times\overline{\mathbb R_-}.$$ To see this, let $w_\lambda=(1-u_1)-\lambda(1-u_2)$.
Then $w_\lambda$ satisfies
the following conditions~:
\begin{itemize}
\item
$-\Delta w_\lambda+am\, w_\lambda>0$ in $\mathbb R\times\mathbb R_-$;
\item
$w_\lambda=(1-\lambda)(1-A)>0$ on $\mathbb R\times\{0\}$.
\end{itemize}
The second property above comes from the fact that both $u_1$ and $u_2$  are equal to $U$
on $\mathbb R\times\overline{\mathbb R_+}$ (cf. Step~1).\\
Now, we denote by~:
$$\lambda_*=\inf\{\lambda \in]0,1]~;\quad (1-u_1)(x)\geq
\lambda(1-u_2)(x)\}.$$
It is then sufficient to prove that $\lambda_*=1$. Suppose by contradiction that $\lambda_*<1$.
Let $w(x)=(1-u_1)(x)-
\lambda_*(1-u_2)(x)$. Then, by the definition of $\lambda_*$, we get~:
\begin{equation}\label{VI-inf=0,R-}
\inf_{x\in\mathbb R\times\mathbb R_-}w(x)=0.
\end{equation}
We claim that we can find a minimizing sequence
$x^n=(x^n_1,x^n_2)\in\mathbb R\times\mathbb R_-$ such that~:
\begin{equation}\label{VI-MinSeq}
\lim_{n\to+\infty}|x_1^n|=+\infty,\quad\lim_{n\to+\infty}x_2^n=b\text{ (for some }b\leq0),\quad
\lim_{n\to+\infty}w(x^n)=0.\end{equation}
Notice that a minimizing sequence can not be bounded, since $u_1,u_2$ satisfy the hypothesis $(H_{\lambda_*}')$
with $\lambda_*\in]0,1[$. Notice that, if $x^n$ is a minimizing sequence then $x^n_2$ should be bounded,
since (cf. Lemma~\ref{VI-limx2}) $\displaystyle\lim_{x_2\to-\infty}w(x_1,x_2)=1-\lambda_*$ uniformly with respect
to $x_1$. So, $x^n_1$ should be unbounded and the existence of a minimizing sequence with Property (\ref{VI-MinSeq})
is clear.\\
We define the function $u_2^n(x_1,x_2)=u_2(x_1+x_2^n,x_2)$. Then $u_2^n$ is a solution of (\ref{VI-limeq}).
We can also extract a subsequence from $u_2^n$ that converges to some function $\widetilde u_2$
in $C_{\rm loc}^2(\overline{\mathbb R\times\mathbb R_+})\cup C_{\rm loc}^2(\overline{\mathbb R\times\mathbb R_-})$
and $\widetilde u_2$ is a solution of (\ref{VI-limeq}). Notice also that
\begin{itemize}
\item $(1-u_1)(0,b)-\lambda_*(1-\widetilde u_2)(0,b)=0$;
\item $u_1$ and $\widetilde u_2$ satisfy the hypothesis $(H_{\lambda_*})$,
\end{itemize}
which is the desired contradiction. Therefore, $\lambda_*=1$.
\hfill$\Box$\\

\section{Asymptotic behavior}

Let $\varepsilon\in]0,\frac1{\sqrt{\lambda_1({\Omega_1})}}[$, then by
(\ref{VI-Min-Max}) and Theorem~\ref{VI-LuPaTh1}, Equation
(\ref{VIGL}) has a unique positive solution $u_\varepsilon$. We
investigate in this section the asymptotic behavior of the solution
$u_\varepsilon$ as $\varepsilon\to0$, proving thus
Theorem~\ref{VI-LuPaTh2}.

\begin{prop}\label{VI-int}(Interior estimate)\ \\
Suppose that the boundaries of $\Omega_1$ and $\Omega$ are of class
 $C^k$
for a given integer $k\geq 1$.  Given $a,m>0$, there exist constants
$\varepsilon_0,\delta,C>0$ such that\footnote{For $k=1$, one is obliged to take $\delta\in]0,\sqrt{am}[$.},
\begin{equation}\label{exp-dec}
\left\|(1-u_\varepsilon)\exp\left(\frac{\delta
t_*(x)}{\varepsilon}\right)\right\|_{H^k({\Omega_1})}
+\left\|u_\varepsilon\exp\left(\frac{\delta
t_*(x)}{\varepsilon}\right)\right\|_{H^k(\Omega_2)}\leq
\frac{C}{\varepsilon^k},\quad\forall\varepsilon\in]0,\varepsilon_0].
\end{equation}
Here $t_*$ is a  function in $C^k(\overline{{\Omega_1}})\cup C^k(\overline{\Omega_2})$  such that
\begin{equation}\label{VI-t*}
0<c\leq\frac{t_*(x)}{{\rm dist}(x,\partial{\Omega_1})}\leq1\quad\text{in }
\overline{\Omega},\quad t_*(x)={\rm
dist}(x,\partial{\Omega_1})\text{ \rm in a neighborhood of }
\partial{\Omega_1},
\end{equation}
and $c\in]0,1[$ is a geometric constant.
\end{prop}
\paragraph{\bf Proof.}
We shall use Agmon type estimates~\cite{Ag}. The technique of Agmon
  estimates is introduced in  the
context of superconductivity by Helffer-Pan~\cite{HePa} (see also
Helffer-Morame~\cite{HeMo3}). The
proof will be  split in two steps, where we first  determine
an estimate in ${\Omega_1}$ and then we determine an
estimate in $\Omega_2$.\\
{\it Step~1. Estimate in ${\Omega_1}$.}\\
We consider~:
$$w_\varepsilon(x)=1-u_\varepsilon(x),\quad c_\varepsilon(x)=(1+u_\varepsilon(x))
u_\varepsilon(x),\quad \quad\forall x\in{\Omega_1}.$$ Using
(\ref{VIGL}), we get,
\begin{equation}\label{VI-subsol}
-\Delta
w_\varepsilon+\frac1{\varepsilon^2}c_\varepsilon(x)w_\varepsilon=0\quad
\text{in }{\Omega_1},\quad -\frac1m\Delta w_\varepsilon+\frac{a}{\varepsilon^2}\,
w_\varepsilon=\frac{a}{\varepsilon^2}\quad\text{in }\Omega_2,
\end{equation}
together with the boundary conditions
$$\mathcal T_{\partial\Omega_1}^{\rm int}(\nu_1\cdot\nabla w_\varepsilon)=\frac1m\mathcal
T_{\partial\Omega_1}^{\rm ext}(\nu_1\cdot\nabla w_\varepsilon),\quad
\mathcal T_{\partial\Omega}^{\rm int}(\nu\cdot\nabla w_\varepsilon)=0.$$
Let $\Phi$ be a Lipschitz function in $\Omega$. An
integration by parts yields the following identity,
\begin{eqnarray}\label{VI-Agm1}
&&\int_{\Omega_1}\left(\left|\nabla\left(e^\Phi
w_\varepsilon\right)\right|^2+\frac1{\varepsilon^2}c_\varepsilon(x)
\left|e^\Phi w_\varepsilon\right|^2\right)\md x\\
&&+\int_{\Omega_2}\left(\frac1m\left|\nabla\left(e^\Phi
w_\varepsilon\right)\right|^2+\frac1{\varepsilon^2}a\left|e^\Phi
w_\varepsilon\right|^2\right)\md x\nonumber\\
&&=\int_{\Omega_1}\left| \,|\nabla\Phi|e^\Phi w_\varepsilon\right|^2\md
x+\int_{\Omega_2}\left(\frac1m\left|\,|\nabla\Phi|e^\Phi
w_\varepsilon\right|^2+\frac{a}{\varepsilon^2}e^{2\Phi}w_\varepsilon\,\md
x\right).\nonumber
\end{eqnarray} 
Lu-Pan \cite[Formula (4.1)]{LuPa96} have proved the following lemma.
\begin{lem}\label{VIlem(4.1)}
Suppose  $u_\varepsilon\in C^2(\overline{\Omega_1})$ be a positive 
solution of $-\Delta
 u_\varepsilon=\frac1{\varepsilon^2}(1-u_\varepsilon^2)u_\varepsilon$ in $\Omega_1$. Then there exist positive constants 
$c_0,k_0,\varepsilon_0$ depending only on ${\Omega_1}$ such that,
\begin{equation}\label{VI-LuPa(4.3)}
\inf_{x\in{\Omega_1},t(x)\geq k_0\varepsilon}u_\varepsilon(x)\geq
c_0,\quad \forall~
\varepsilon\in]0,\varepsilon_0]. \end{equation}
Here $t$ is defined by (\ref{VI-t(x)}).\end{lem}
We emphasize that no necessary hypothesis is needed concerning 
the boundary condition in Lemma~\ref{VIlem(4.1)}.
For the convenience of the reader, we shall reproduce  the proof
of Lemma~\ref{VIlem(4.1)} in Appendix~\ref{Appendix-Lemma}.\\
We come back to the proof of Proposition \ref{VI-int}. By the lemma, we get 
\begin{equation}\label{VI-c-epsilon>c0}
c_\varepsilon(x)\geq c_0,\quad \forall x\in\Omega_1\text{
  s.t. }t(x)\geq
  k_0\varepsilon 
,\quad\forall~\varepsilon\in]0,\varepsilon_0].\end{equation}
We choose the function $\Phi$ in the following form,
$$\Phi=\frac{\delta}{\varepsilon}\phi,$$
where $\delta>0$ is to be determined and
$$
\phi(x)= \left\{
\begin{array}{l}
t(x);\quad \text{if }t(x)\geq k_0\varepsilon,\\
k_0\varepsilon;\quad\text{if }t(x)\leq k_0\varepsilon.
\end{array}\right.
$$
Coming back to
(\ref{VI-Agm1}) and (\ref{VI-c-epsilon>c0}), we obtain the following
estimate,
\begin{eqnarray}\label{VI-Agm2}
&&\int_{\Omega_1}\left(\varepsilon^2\left|\nabla\left(\exp\left(\frac{\delta\phi}{\varepsilon}\right)
w_\varepsilon\right)\right|^2+(c_0-\delta^2)
\left|\exp\left(\frac{\delta\phi}{\varepsilon}\right) w_\varepsilon\right|^2\right)\md x\\
&&+\int_{\Omega_2}\left(\varepsilon^2\frac1m\left|\nabla\left(
\exp\left(\frac{\delta\phi}{\varepsilon}\right)
w_\varepsilon\right)\right|^2+a\left|\exp\left(\frac{\delta\phi}{\varepsilon}\right)
w_\varepsilon\right|^2\right)\md x\nonumber \\
&&\leq a\int_{\Omega_2} e^{2\delta\phi/\varepsilon} w_\varepsilon\,\md
x.\nonumber
\end{eqnarray}
Upon taking $\delta\in]0,\sqrt{c_0})[$, the above estimate reads as,
\begin{equation}\label{VI-contH1}
\left\|\exp\left(\frac{\delta\phi}{\varepsilon}\right)
w_\varepsilon\right\|_{L^2(\Omega)}+
\varepsilon\left\|\nabla\left(\exp\left(\frac{\delta\phi}{\varepsilon}\right)
w_\varepsilon\right)\right\|_{H^1(\Omega)}\leq C,
\end{equation}
where the constant $C$ depends on $a,m,\Omega_1$ and ${\Omega_2}$.
We emphasize here that the function $t$ is negative in $\Omega_2$ so
that $\phi(x)=k_0\varepsilon$.\\
Let $t_*$ be verifying (\ref{VI-t*}). We
can select $t_*$ in the following way,
\begin{eqnarray*}
&&t_*(x)=|t(x)|\quad\text{in }{\Omega_1}(k_0/2),\\
&&t_*(x)=\frac{k_0}2+\chi\left(\frac{t(x)}{k_0}\right)
\left(|t(x)|-\frac{k_0}2\right)\quad\text{ in }\mathbb
R^2\setminus\overline{{\Omega_1}(k_0/2)}, \end{eqnarray*} where $\chi$
is a cut-off function that verifies~:
\begin{equation}\label{VI-chi}
0\leq\chi\leq1,\quad \chi\equiv1\text{ in }]-\frac12,\frac12[,\quad
{\rm supp}\,\chi\subset[-1,1].
\end{equation}
Noticing that $t_*(x)\leq t(x)\leq\phi(x)$ in ${\Omega_1}$, we deduce
from (\ref{VI-contH1}) the following control,
\begin{equation}\label{VI-contL2}
\left\|\exp\left(\frac{\delta t_*(x)}{\varepsilon}\right)
w_\varepsilon\right\|_{L^2({\Omega_1})}
+\varepsilon\left\|\nabla\left(\exp\left(\frac{\delta
t_*(x)}{\varepsilon}\right)w_\varepsilon\right)\right\|_{L^2({\Omega_1})}
\leq C.
\end{equation}
To derive higher order Sobolev estimates, we look at the PDE
satisfied by $\exp\left(\frac{\delta t_*(x)}{\varepsilon}\right)
w_\varepsilon$. Let us define the following function,
$$v_\varepsilon(x)=\left\{
\begin{array}{l}
\exp\left(\frac{\delta t_*(x)}{\varepsilon}\right)
w_\varepsilon,\quad\text{in }{\Omega_1},\\
\\
\exp\left(-\frac{\delta
t_*(x)}{\varepsilon}\right)w_\varepsilon,\quad\text{in }\Omega_2.
\end{array}\right.
$$
Then $v_\varepsilon$  is a weak solution of the following equation,
\begin{equation}\label{VIGL-v}
\left\{
\begin{array}{l}
-\Delta v_\varepsilon=f_{\varepsilon,1}\quad\text{in }{\Omega_1},\\
\\
-\frac1m\Delta v_\varepsilon=f_{\varepsilon,2}\quad\text{in }\Omega_2\cap{\Omega_1}(k_0),\\
\\
\mathcal T_{\partial{\Omega_1}}^{\rm int}(\nu\cdot \nabla v_\varepsilon)=\frac1m\mathcal
T_{\partial{\Omega_1}}^{\rm ext}(\nu\cdot\nabla v_\varepsilon)\quad\text{on
}\partial{\Omega_1}.
\end{array}
\right.
\end{equation}
Here, the set ${\Omega_1}(k_0)$ is defined by (\ref{VI-Om-t}), and the
functions $f_{\varepsilon,1},f_{\varepsilon,2}$ are given by,
\begin{eqnarray*}
&&f_{\varepsilon,1}=\frac1{\varepsilon^2}(1-w_\varepsilon)(2-w_\varepsilon)v_\varepsilon
-2\frac\delta\varepsilon\nabla t_*\exp\left(\frac{\delta
t_*(x)}{\varepsilon}\right) \cdot\nabla w_\varepsilon
-\frac{\delta}{\varepsilon}\left(\Delta
t_*+\frac{\delta}{\varepsilon}|\nabla t_*|^2\right)v_\varepsilon,\\
&&f_{\varepsilon,2}=-\frac a{\varepsilon^2}v_\varepsilon+\frac1m\left(\frac{\delta}{\varepsilon}\right)
\left(2\nabla t_*\exp\left(\frac{-\delta t_*(x)}{\varepsilon}\right)
\cdot\nabla w_\varepsilon +\frac{\delta}{\varepsilon}\left(\Delta
t_*+\frac{\delta}{\varepsilon}|\nabla
t_*|^2\right)v_\varepsilon\right).
\end{eqnarray*}
Using Theorem~\ref{VI-regKa} together with (\ref{VI-contL2}), we
get,
$$\|v_\varepsilon\|_{H^2({\Omega_1})}\leq C\varepsilon^{-2}.$$
Applying Theorem~\ref{VI-regKa} recursively,  we get for any integer
$k\geq1$,
\begin{equation}\label{VI-estOm}
\left\|\exp\left(\frac{\delta t_*(x)}{\varepsilon}\right)
w_\varepsilon\right\|_{H^k({\Omega_1})}\leq C\varepsilon^{-k}.
\end{equation}
{\it Step~2. Estimate in $\Omega_2$.}\\
We apply the same argument as in Step~1 (which is actually simpler
in this case since the equation satisfied by $u_\varepsilon$ in
$\Omega_2$ is linear), and only sketch the main points of the
proof. Let $\Phi$ be again a Lipschitz function. An integration by
parts yields the following identity,
\begin{equation}\label{VI-Agm1d}
\mathcal G_0\left(e^\Phi u_\varepsilon\right)=
\left\|\,|\nabla\Phi|e^\Phi
u_\varepsilon\right\|_{L^2({\Omega_1})}^2+\frac1m\left\|\,|\nabla\Phi|e^\Phi
u_\varepsilon\right\|_{L^2(\Omega_2)}^2.
\end{equation}
Similarly as Step~1, we choose $\Phi$ in the following form,
$$\Phi(x)=\frac{\delta}{\varepsilon}t_*(x),\quad \text{in
}\Omega_2,\quad \Phi(x)=0\quad\text{in } {\Omega_1},
$$
with $\delta>0$. By taking\footnote{It is here that we observe the dependence of $\delta$ on $am$.}  $\delta\in]0,\sqrt{ma}[$, we get from
(\ref{VI-Agm1d}) the following control on the $H^1$-norm,
$$\left\|\exp\left(\frac{\delta
t_*(x)}{\varepsilon}\right)u_\varepsilon\right\|_{L^2(\Omega_2)}+\varepsilon
\left\|\nabla\left(\exp\left(\frac{\delta
t_*(x)}{\varepsilon}\right)u_\varepsilon\right)\right\|_{L^2(\Omega_2)}\leq
C,$$ for some constant $C>0$ depending only on $a,m,\Omega_1$ and ${\Omega_2}$.
Using Theorem~\ref{VI-regKa}, we can derive higher Sobolev
estimates. Actually, for any integer $k\geq1$, we can find a
constant $C>0$ such that,
\begin{equation}\label{VI-estOmd}
\left\|\exp\left(\frac{\delta
t_*(x)}{\varepsilon}\right)u_\varepsilon\right\|_{H^k(\Omega_2)}\leq
C\varepsilon^{-k}.
\end{equation}
Combined with (\ref{VI-estOm}),  the above estimate permits us to
deduce (\ref{exp-dec}) and thus to prove
Proposition~\ref{VI-int}.\hfill$\Box$\\

\begin{rem}\label{VI-comm-LuPa}
The argument given in \cite[(4.2)-(4.3)]{LuPa96} permits us also to prove
an exponential decay of $1-u_\varepsilon$ in ${\Omega_1}$.
The proof of \cite{LuPa96} relies in part on a result of
Fife~\cite[p.~230]{F}. We
have used here Agmon type estimates~\cite{Ag}.
\end{rem}

\begin{prop}\label{VI-bnd}(Boundary estimate)\ \\
Let $\Gamma$ be a connected component of $\partial{\Omega_1}$. Given
$R>0$, there exists a constant $\varepsilon_0$ depending only on
$R,a,m$ and ${\Omega_1}$ such that, if
$\varepsilon\in]0,\varepsilon_0]$ and $\varepsilon\to0$, then,
\begin{equation}\label{VI-bnd-eq}
\left\|u_\varepsilon(x)-U\left(\frac{t(x)}{\varepsilon}\right)\right\|_{L^\infty(\Gamma(\varepsilon
R))}=o(1).
\end{equation}
Here, the functions $U,t$  are defined respectively by
(\ref{VI-Sol}) and (\ref{VI-t(x)}), and for a given $\delta>0$, the
set $\Gamma(\delta)\subset\Omega$ is defined by,
$$\Gamma(\delta)=\{x\in \Omega~:\quad {\rm
dist}(x,\Gamma)\leq \delta\}.$$
\end{prop}
\paragraph{\bf Proof.}
We  work with the
$(s,t)$-coordinates defined by
(\ref{Phi(s,t)}). We can in addition assume that~:
\begin{equation}\label{VI-gam}
\Gamma=\{x\in\partial{\Omega_1}~:\quad t(x)=0,\quad
-\frac{|\Gamma|}2\leq s(x)\leq \frac{|\Gamma|}2\,\}.
\end{equation}
Let $\widetilde u_\varepsilon$ be the function assigned to
$u_\varepsilon$ by (\ref{VI-Tu}). Notice that, thanks to
(\ref{VI-TDel}), $\widetilde u_\varepsilon$ satisfies the following
equation~:
\begin{equation}\label{VI-eq(s,t)}
\left\{
\begin{array}{l}
-\widetilde\Delta\,\widetilde u_\varepsilon
=\frac1{\varepsilon^2}(1-\widetilde u_\varepsilon^2)\widetilde
u_\varepsilon,\quad \text{for }0<t<t_0\text{ and }-\frac{|\Gamma|}2< s< \frac{|\Gamma|}2,\\
\\
-\widetilde \Delta\,\widetilde u_\varepsilon+\frac{am}{\varepsilon^2}\,\widetilde
u_\varepsilon=0,\quad\text{for }-t_0<t<0\text{ and
}-\frac{|\Gamma|}2< s< \frac{|\Gamma|}2,\\
\\
\displaystyle\frac{\partial\widetilde u_\varepsilon}{\partial
t}(\cdot,0_+)=\displaystyle\frac1m\displaystyle\frac{\partial\widetilde
u_\varepsilon}{\partial t}(\cdot,0_-),\quad \text{for }t=0.
\end{array}
\right.
\end{equation}
We define the following rescaled function~:
\begin{equation}\label{VI-rescFunc}
\widetilde v_\varepsilon(s,t)=\widetilde u_\varepsilon(\varepsilon
s,\varepsilon t),
\end{equation} then, thanks to (\ref{VI-eq(s,t)}),
$\widetilde v_\varepsilon$ satisfies the following equation,
\begin{equation}\label{VI-limeq(s,t)}
\left\{
\begin{array}{l}
-\Delta_\varepsilon\, \widetilde v_\varepsilon =(1-\widetilde
v_\varepsilon^2)\widetilde v_\varepsilon,\quad
\text{for }0<t<\frac{t_0}{\varepsilon}\text{ and }-\frac{|\Gamma|}{2\varepsilon}< s< \frac{|\Gamma|}{2\varepsilon},\\
\\
-\Delta_\varepsilon\,\widetilde v_\varepsilon+am\,\widetilde v_\varepsilon=0,\quad\text{for
}-\frac{t_0}{\varepsilon}<t<0\text{ and
}-\frac{|\Gamma|}{2\varepsilon}< s< \frac{|\Gamma|}{2\varepsilon},\\
\\
\displaystyle\frac{\partial\widetilde v_\varepsilon}{\partial
t}(\cdot,0_+)=\displaystyle\frac1m\displaystyle\frac{\partial\widetilde
v_\varepsilon}{\partial t}(\cdot,0_-),\quad \text{for }t=0.
\end{array}
\right.
\end{equation}
Here the operator $\Delta_\varepsilon$ is given by~:
$$\Delta_\varepsilon=\left(1-\varepsilon t\kappa_{\rm r}(\varepsilon
s)\right)^{-2}\partial_s^2+\partial_t^2 + \frac{\varepsilon^2
t\kappa_{\rm r}'(\varepsilon s)}{\left(1-\varepsilon t\kappa_{\rm
r}(\varepsilon s)\right)^3}\partial_s-\frac{\varepsilon \kappa_{\rm
r}(\varepsilon s)}{\left(1-\varepsilon t\kappa_{\rm r}(\varepsilon
s)\right)}\partial_t.$$ Let $K\subset\mathbb R^2$ be a compact set,
then there exists $\varepsilon_0(K)>0$ such that, for
$\varepsilon\in]0,\varepsilon_0(K)]$, $K\subset\{|t|\leq
t_0/\varepsilon,|s|\leq |\Gamma|/(2\varepsilon)\}$.\\
By Theorem~\ref{VI-regKa}, there exists a constant $C(K)>0$ such that,
$$\|\widetilde v_\varepsilon\|_{H^4(K_+)}+\|\widetilde v_\varepsilon\|_{H^4(K_-)}\leq
C(K),\quad \forall \varepsilon\in]0,\varepsilon_0(K)],$$ where
$K_+=K\cap\{t>0\}$ and $K_-=K\cap\{t<0\}$.\\
By the Sobolev Imbedding
Theorem, we get,
$$\|\widetilde
v_\varepsilon\|_{C^{2,\alpha}(\overline{K_+})}+\|\widetilde v_\varepsilon\|_{C^{2,\alpha}(\overline{K_-})}\leq \widetilde
C(\alpha,K),\quad\forall~\alpha\in]0,1[,\quad\forall~\varepsilon\in]0,\varepsilon_0(K)].$$ Therefore,
by passing to a subsequence, we may assume that
$$\widetilde v_\varepsilon\to v\text{ in }C_{\rm loc}^2(\overline{\mathbb R\times\mathbb R_+})\text{ and in } C_{\rm
loc}^2(\overline{\mathbb R\times\mathbb R_-}).$$ Notice that $0\leq
v\leq 1$, $v$  is a solution of (\ref{VI-limeq}) and, by (\ref{VI-LuPa(4.3)}) and (\ref{VI-rescFunc}),
$$\exists\, k_0,c_0>0,\quad v(0,k_0)\geq c_0>0.$$
Therefore, by Theorem~\ref{VI-sol-modeq}, we get that $v=U(t)$,
where $U$ is the one-dimensional solution. Thus given $R>0$, we have,
\begin{equation}\label{VI-PtwConv}
\lim_{\varepsilon\to0}\left\|\widetilde v_\varepsilon(s,t)-U(t)\right\|_{W^{2,\infty}(\{|s|\leq
R,|t|\leq R\})}=0.\end{equation} Coming back to the definition of
$v_\varepsilon$, the above limit reads as,
$$\lim_{\varepsilon\to0}\left\|\widetilde
u_\varepsilon-U\left(\frac{t}{\varepsilon}\right)\right\|_{L^\infty(\{|s|\leq
\varepsilon R,|t|\leq \varepsilon R\})}=0,$$ and this achieves the
proof of the proposition.\hfill$\Box$\\

\noindent\paragraph{\bf Proof of Theorem~\ref{VI-LuPaTh2}.}\ \\
{\it Proof of (\ref{VIu-lim0}).}\\
This is a consequence of Proposition~\ref{VI-int} and of the Sobolev
Imbedding Theorem.\\
{\it Proof of (\ref{VI-bndlay-S}).}\\
Let $w_\varepsilon(x)=u_\varepsilon(x)-U(t(x)/\varepsilon)$. Let
$x_\varepsilon\in\overline\Omega$ be a point of maximum of
$w_\varepsilon$,
$$w_\varepsilon(x_\varepsilon)=\|w_\varepsilon\|_{L^\infty(\overline\Omega)}.$$
If $t(x_\varepsilon)/\varepsilon$ is bounded, we get by
Proposition~\ref{VI-bnd} that
$$\lim_{\varepsilon\to0}w_\varepsilon(x_\varepsilon)=0.$$
Otherwise, if $\displaystyle\lim_{\varepsilon\to0}
|t(x_\varepsilon)/\varepsilon|=+\infty$, then we get by Proposition~\ref{VI-int},
$$\lim_{\varepsilon\to0}w_\varepsilon(x_\varepsilon)=0.$$
Therefore, $w_\varepsilon\to0$ uniformly in $\overline\Omega$.\\

\section{Energy estimate}\label{Sec-Enest}

\subsection{A one-dimensional variational problem}
The proof of Theorem~\ref{VIen} relies on an auxiliary result
concerning a one-dimensional variational problem. Let us introduce the
space~:
\begin{equation}\label{V-1Dspace}
\mathcal H=\{u\in L^2_{\rm loc}(\mathbb R)~:~u'\in L^2(\mathbb R),~
1-|u|\in L^2(\mathbb R_+),~u\in L^2(\mathbb R_-)\}.
\end{equation}
Our objective is to minimize the functional~:
\begin{equation}\label{V-1DFunctional}
\mathcal F(u)=\int_{0}^{+\infty}\left(|u'(t)|^2\,\md
t+\frac12(1-u^2(t))^2\right)\md t
+\int_{-\infty}^0\left(\frac1m|u'(t)|^2+a\,u^2(t)\right)\md t,
\end{equation}
over the space $\mathcal H$. Since the space $\mathcal H$ is 
continuously embedded in $L^\infty(\mathbb R)$, the functional
$\mathcal F$ is well defined on $\mathcal H$. 

\begin{thm}\label{V-thm-1Dproblem}
The function $U$ (introduced in (\ref{VI-Sol})) minimizes the energy
functional $\mathcal F$ over the space $\mathcal H$. Moreover, the
only minimizers in $\mathcal H$ are $\pm U$.
\end{thm}
\paragraph{\bf Proof.}  
Starting from a minimizing sequence, it is standard (cf. e.g. \cite{GJ})
to prove the existence of a minimizer of $\mathcal F$.\\
Now, if $v$ minimizes $\mathcal F$ in $\mathcal H$, then so is
$|v|$. Hence, it is sufficient to look for minimizers in the class
$$\mathcal C=\{u\in L^\infty(\mathbb R)~:~u\geq0,\quad u_{|_{\mathbb
R\times\mathbb R_\pm}}\in C^2(\mathbb R\times\mathbb R_\pm)\}.$$
It results from Theorem~\ref{VI-sol-modeq}  that the Euler-Lagrange
equation associated with the functional $\mathcal F$ admits a unique
solution in $\mathcal C$, given by the function $U$. Hence, the function $U$
minimizes $\mathcal F$ and $\pm U$ are  the only minimizers.\hfill$\Box$\\

Given $a,m>0$, let us introduce the two parameters~:
\begin{equation}
c_1(a,m)=\int_0^{+\infty}\left(|U'(t)|^2+\frac12(1-U^2(t))^2\right)\md
  t
+\int_{-\infty}^0\left(\frac1m|U'(t)|^2+a\,U^2(t)\right)\md
  t,\label{VI-c1}\end{equation}
\begin{equation}
c_2(a,m)=\int_0^{+\infty}\left(|U'(t)|^2+\frac12(1-U^2(t))^2\right)\,t\md
  t
+\int_{-\infty}^0\left(\frac1m|U'(t)|^2+a\,U^2(t)\right)\,t\md
  t.\label{VI-c2}
\end{equation}
An easy computation gives~:
\begin{eqnarray}\label{VI-c1+}
&&c_1(a,m)=\frac{4\sqrt{2}(3\beta+1)}{3(\beta+1)^3}
+\frac12\sqrt{\frac am}\left(1+\frac1{m}\right)A^2,\\
\label{VI-c2+}
&&c_2(a,m)=\frac43\left[\ln\left(\frac{1+\beta}\beta\right)-
\frac\beta{(1+\beta)^2}\right]+\frac{A^2}4\left(1+\frac1m\right),
\end{eqnarray}
where the constants $\beta$ and $A$ are introduced in
(\ref{VI-l,bet}).

\subsection{Upper bound}
Given $a,m>0$, we establish the existence of positive constants
$\delta_0$ and $\varepsilon_0$ such that, for all $\varepsilon\in]0,\varepsilon_0]$,
\begin{equation}\label{VI-ub}
C_0(\varepsilon)\leq 
c_1(a,m)\frac{|\partial\Omega_1|}\varepsilon-c_2(a,m)
\int_{\partial\Omega_1}\kappa_{\rm r}(s)\,\md s
+\mathcal O\left(\exp\left(-\frac{\delta_0}{\varepsilon}\right)\right).
\end{equation}
Here $C_0(\varepsilon)$ is defined in
(\ref{VI-infEn}).\\
Let us define the following function~:
$$v_\varepsilon(x)=U\left(\frac{t(x)}\varepsilon\right)\quad
(x\in\Omega),$$
where the functions $U$ and $t$ has been introduced in (\ref{VI-Sol}) and
(\ref{VI-t}) respectively.
Let us take $t_0>0$ sufficiently small such that 
the coordinate transformation (\ref{Phi(s,t)}) is well defined in 
${\Omega_1}(t_0)$. Here we recall the definition of $\Omega_1(t_0)$
given 
in (\ref{VI-Om-t}).\\
Notice that $v_\varepsilon\in H^1(\Omega)$, and
$$\nabla
v_\varepsilon(x)=\frac1\varepsilon U'\left(\frac{t(x)}\varepsilon\right)\,\nabla
t(x),\quad\forall~x\in\Omega.$$
Let us compute the energy $\mathcal G_0(v_\varepsilon)$. Notice that,
due to the expression of $v_\varepsilon$ and $U$, and since $|\nabla
t(x)|=1$ (cf. Subsection~\ref{VI-BndCord}),
  we get a constant $\delta_1>0$ such that,
\begin{equation}\label{VI-splitting}
\mathcal G_0(v_\varepsilon)=\mathcal G_0(v_\varepsilon,\Omega_1(t_0))
+\mathcal O\left(\exp\left(-\frac{
\delta_1}{\varepsilon}\right)\right),
\end{equation}
where 
\begin{eqnarray*}
&&\hskip-1.5cm\mathcal G_0(v_\varepsilon,\Omega_1(t_0))
=\int_{\Omega_1\cap\Omega_1(t_0)}\left(|\nabla v_\varepsilon|^2
+\frac1{2\varepsilon^2}(1-v_\varepsilon^2)^2\right)\,\md x\\
&&\hskip4cm+\int_{\Omega_2\cap\Omega_1(t_0)}\left(
\frac1m|\nabla
v_\varepsilon|^2+\frac{a}{\varepsilon^2}v_\varepsilon^2\right)\,\md x.
\end{eqnarray*}
We express the energy interms of $(s,t)$ boundary coordinates which
are valid in $\Omega_1(t_0)$. It is a result  of (\ref{VI-WL2}) that,  
\begin{eqnarray*}
\mathcal G_0(v_\varepsilon,\Omega_1(t_0))
&=&\frac1{\varepsilon^2}\int_{-|\partial\Omega_1|/2}^{|\partial\Omega_1|/2}
\int_0^{t_0}\left(\left|U'\left(\frac
t\varepsilon\right)\right|^2+\frac1{2}
\left(1-U^2\left(\frac t\varepsilon\right)\right)^2\right){\rm a}(s,t)\md
t\md s\\
&&+\frac1{\varepsilon^2}\int_{-|\partial\Omega_1|/2}^{|\partial\Omega_1|/2}
\int_{-t_0}^{0}\left(\frac1m\left|U'\left(\frac
t\varepsilon\right)\right|^2+a\,
U^2\left(\frac t\varepsilon\right)\right){\rm a}(s,t)\,\md t\md s,
\end{eqnarray*}
where we recall that ${\rm a}(s,t)=1-t\kappa_{\rm r}(s)$.\\
Performing the scaling $s=\varepsilon\widetilde s$ and
$t=\varepsilon\widetilde t$, we get (we remove the tildes for simplicity)~:
\begin{eqnarray*}
\mathcal G_0(v_\varepsilon,\Omega_1(t_0))&=&
\int_{-|\partial\Omega_1|/2\varepsilon}^{|\partial\Omega_1|/2\varepsilon}
\int_{0}^{t_0/\varepsilon}\left(\left|U'(
t)\right|^2+\frac12
\left(1-U^2(t)\right)^2\right)(1-\varepsilon t\kappa_{\rm
  r}(\varepsilon s))\md
t\md s\\
&&+\int_{-|\partial\Omega_1|/2\varepsilon}^{|\partial\Omega_1|/2\varepsilon}
\int_{-t_0/\varepsilon}^0\left(\frac1m\left|U'(
t)\right|^2+a
U^2( t)\right)
(1-\varepsilon t\kappa_{\rm r}(\varepsilon s))\,\md t\md s.
\end{eqnarray*}
Using the exponential decay of $U$ and $U'$ at $\pm\infty$, we obtain a
constant $\delta_1>0$ such that,
$$\mathcal G_0(v_\varepsilon,\Omega_1(t_0))=
c_1(a,m)\frac{|\partial\Omega_1|}\varepsilon-c_2(a,m)\int_{\partial\Omega_1}
\kappa_{\rm r}(s)
+\mathcal O\left(\exp\left(-\frac{\delta_2}\varepsilon
\right)\right),$$
where $c_1(a,m)$ and $c_2(a,m)$ are introduced in (\ref{VI-c1}) and 
(\ref{VI-c2}) respectively.\\
Coming back to (\ref{VI-splitting}) and 
recalling that $C_0(\varepsilon)\leq \mathcal G_0(v_\varepsilon)$, 
we  get the
upper bound announced in (\ref{VI-ub}).\\

\subsection{Lower bound} 
We establish the following lower bound,
\begin{equation}\label{VI-imp-LB}
\mathcal G_0(u_\varepsilon)\geq
c_1(a,m)\frac{|\partial\Omega_1|}\varepsilon
-c_2(a,m)\int_{\partial\Omega_1}\kappa_{\rm r}(s)\,\md s+o(1),\quad
(\varepsilon\to0).
\end{equation}
Proposition~\ref{VI-int} will reduce the analysis to a thin region
$\Omega_1(\varepsilon^\ell)$, where $\ell\in]0,1[$ can be chosen
arbitrarly.\\
Actually, we write,
\begin{eqnarray}\label{VI-redOm(t0)}
&&\mathcal G_0(u_\varepsilon)=
\mathcal G_0\left(u_\varepsilon,\Omega_1(\varepsilon^\ell)\right)
+\mathcal G_0
\left(u_\varepsilon,\Omega\setminus\Omega_1(\varepsilon^\ell)\right),
\end{eqnarray}
where, for a given $\mathcal U\subset \Omega$,
$$\mathcal G_0(u_\varepsilon,\mathcal U)=
\int_{{\Omega_1}\cap \mathcal U}\left(
|\nabla
u_\varepsilon|^2+\frac1{2\varepsilon^2}(1-|u_\varepsilon|^2)^2\right)\md
x+\int_{\Omega_2\cap \mathcal U}
\left(\frac1m|\nabla
u_\varepsilon|^2+\frac{a}{\varepsilon^2}|u_\varepsilon|^2\right)\md
x.$$
We observe that, by Proposition~\ref{VI-int}, the last term in 
(\ref{VI-LuPaTh2}), $\mathcal
G_0(u_\varepsilon,\Omega\setminus\Omega_1(\varepsilon^\ell))$,  
is exponentially small as $\varepsilon\to0$.\\
It is then sufficient to look for a lower bound of the reduced energy
$\mathcal G_0(u_\varepsilon,\Omega_1(\varepsilon^\ell))$. Let us
express this energy in the $(s,t)$ coordinates introduced in 
Section~\ref{VI-BndCord}. Actually, we write,
\begin{eqnarray}\label{V-Pr-EnH=0}
&&\\
&&\hskip-0.5cm\mathcal G_0\left(u_\varepsilon,\Omega_1(\varepsilon^\ell)\right)=
\int_{-|\partial \Omega_1|/2}^{|\partial \Omega_1|/2}
\int_{0}
^{\varepsilon^\ell}\left(|\partial_tu_\varepsilon|^2+{\rm
  a}^{-2}|\partial_su_\varepsilon|^2+
\frac1{2\varepsilon^2}
(1-u_\varepsilon^2)^2\right)\,{\rm a}(s,t)\,\md t\md s\nonumber\\
&&\hskip1.5cm+\int_{-|\partial \Omega_1|/2}^{|\partial \Omega_1|/2}
\int_{-\varepsilon^\ell}^{0}
\left(|\partial_tu_\varepsilon|^2+ {\rm a}^{-2}|\partial_su_\varepsilon|^2
+\frac{a}{\varepsilon^2}u_\varepsilon^2\right)\,
{\rm a}(s,t)\,\md t\md s,\nonumber
\end{eqnarray}
where ${\rm a}(s,t)=1-t\kappa_{\rm r}(s,t)$ is the Jacobian of the
coordinate transformation.\\
We get immediatly the following  simple `lower bound' decomposition of 
(\ref{V-Pr-EnH=0})~:
\begin{equation}\label{VI-imp-LB'}
\mathcal G_0\left(u_\varepsilon,\Omega_1(\varepsilon^\ell)\right)\geq
\int_{-|\partial \Omega_1|/2}^{|\partial \Omega_1|/2}
\left(\mathcal F_\varepsilon(u_\varepsilon)
-\mathcal R_\varepsilon(u_\varepsilon)\kappa_{\rm r}(s)\right)\,\md s,
\end{equation}
where
\begin{equation}\label{VI-LB-F}
\mathcal F_\varepsilon(u_\varepsilon)=
\int_0^{\varepsilon^\ell}\left(|\partial_tu_\varepsilon|^2+
\frac1{2\varepsilon^2}(1-u_\varepsilon^2)^2\right)\md t+
\int_{-\varepsilon^\ell}^0\left(\frac1m|\partial_tu_\varepsilon|^2+\frac{a}
{\varepsilon^2}u_\varepsilon^2\right)\,\md t,
\end{equation}
\begin{equation}\label{VI-LB-R}
\mathcal R_\varepsilon(u_\varepsilon)=
\int_0^{\varepsilon^\ell}\left(|\partial_tu_\varepsilon|^2+
\frac1{2\varepsilon^2}(1-u_\varepsilon^2)^2\right)\,t\md t+
\int_{-\varepsilon^\ell}^0\left(\frac1m|\partial_tu_\varepsilon|^2+\frac{a}
{\varepsilon^2}u_\varepsilon^2\right)\,t\md t.
\end{equation}
Let us define the rescaled function
$$\widetilde v_\varepsilon(\widetilde s,\widetilde t)=
u_\varepsilon(\varepsilon 
\widetilde s,\varepsilon\widetilde t),\quad
-\frac{|\partial\Omega_1|}{2\varepsilon}\leq \widetilde s\leq
\frac{|\partial\Omega_1|}{2\varepsilon},~
-\varepsilon^{\ell-1}\leq \widetilde t\leq \varepsilon^{\ell-1},$$
and we extend it by continuity to $\mathbb R^2$.\\
Then, by applying again Proposition~\ref{VI-int}, we get positive
constants
$C$ and $\delta_0$ such that,
$$\left|\mathcal F_\varepsilon(u_\varepsilon)-\frac1{\varepsilon}
\mathcal F(\widetilde v_\varepsilon)\right|\leq
\frac{C}{\varepsilon}\exp\left(-\frac{\delta_0}{\varepsilon^{\ell-1}}\right),
$$
where $\mathcal F$ is the functional introduced in
(\ref{V-1DFunctional}). As the function $\widetilde t\mapsto
\widetilde v_\varepsilon(\widetilde s,\widetilde t)$ 
is in the space $\mathcal H$
introduced in (\ref{V-1Dspace}), we deduce by
Theorem~\ref{V-thm-1Dproblem} and upon recalling the
definition of $c_1(a,m)$ in (\ref{VI-c1})~:
$$\mathcal F_\varepsilon(u_\varepsilon)\geq
\frac{c_1(a,m)}\varepsilon-\frac{C}{\varepsilon}
\exp\left(-\frac{\delta_0}{\varepsilon^{\ell-1}}\right).$$
Substituting in (\ref{VI-imp-LB'}), we get,
\begin{equation}\label{VI-imp-LB''}
\mathcal G_0(u_\varepsilon,\Omega_1(\varepsilon^{\ell-1}))
\geq \int_{-|\partial\Omega_1|/2\varepsilon}^
{|\partial\Omega|/2\varepsilon}\left(c_1(a,m)+\varepsilon
\widetilde{\mathcal
  R}_\varepsilon\kappa_{\rm r}(\varepsilon \widetilde s)\right)\,\md
\widetilde s-\frac{\widetilde C}{\varepsilon^2}
\exp\left(-\frac{\delta_0}{\varepsilon^{\ell-1}}\right),
\end{equation}
where
$$\widetilde{\mathcal R}_\varepsilon
=
\int_0^{\varepsilon^{\ell-1}}\left(|\partial_t\widetilde v_\varepsilon|^2+
\frac12(1-\widetilde v_\varepsilon^2)^2\right)\,t\md t+
\int_{-\varepsilon^\ell}^0
\left(\frac1m|\partial_t\widetilde v_\varepsilon|^2+a\widetilde 
v_\varepsilon^2\right)\,t\md
t.$$
It is then sufficient to prove that $\widetilde
R_\varepsilon\geq c_2(a,m)+o(1)$, where $c_2(a,m)$ has been introduced in 
(\ref{VI-c2}).\\
Actually, by (\ref{VI-PtwConv}), $\widetilde v_\varepsilon$  
converges
pointwise to the function $U$ which has been introduced in
(\ref{VI-Sol})
(a similar convergence result holds for the derivatives). Moreover,
since $u_\varepsilon$ minimizes $\mathcal G_0$, we have by the upper bound
(\ref{VI-ub}) that $\widetilde R_\varepsilon$ is bounded as
$\varepsilon\to0$. Now, applying Fatou's lemma, we get
$$\widetilde R_\varepsilon \geq 
\int_0^{+\infty}\left(|U'(t)|^2+\frac12(1-U^2(t))^2\right)\,t\md t
+\int_{-\infty}^0\left(\frac1m|U'(t)|^2+a\,U^2(t)\right)\,t\md t+o(1).$$
Recalling the definition of $c_2(a,m)$ in (\ref{VI-c2}), we are in a
position to deduce from (\ref{VI-imp-LB'}),
$$\mathcal G_0(u_\varepsilon,\Omega_1(\varepsilon^{\ell-1}))\geq
c_1(a,m)\frac{|\partial\Omega_1|}\varepsilon-c_2(a,m)
\int_{\partial\Omega_1}\kappa_{\rm r}(s)\,\md s+o(1).$$
This is sufficient, upon recalling the remark concerning
(\ref{VI-redOm(t0)}), to achieve the proof of the lower bound
announced in (\ref{VI-imp-LB}).
\subsection{Proof of Theorem~\ref{VIen}}
It is sufficient to combine  the upper bound (\ref{VI-ub}) with the
lower bound in (\ref{VI-imp-LB}).

\section{Concluding remarks}
\subsection{Link with the physical literature (The breakdown field).}\ \\
Let us come back to the physical interpretation of Equation
(\ref{VIGL}). It is supposed that ${\Omega_1}$ is occupied by a
superconducting material and $\Omega_2$ by a normal
metal. The function $u_\varepsilon^2$ measures the density of the
superconducting electrons (Cooper pairs) so that 
$u_\varepsilon\approx0$ corresponds to a non-superconducting
region.\\
In the particular case when $\Omega_1=\mathbb R\times\mathbb R_+$
and $\Omega_2=\mathbb R\times\mathbb R_-$, we obtained that the
solution satisfies the de\,Gennes boundary condition
(\ref{VI-deGebndcn}) on the boundary of $\Omega_1$. In
(\ref{VI-deGebndcn}), the parameter $\gamma$ is  called {\it the
de\,Gennes parameter}. One also defines the  {\it extrapolation
length} by $b:=\frac1{\gamma}$ which is given now by~:
\begin{equation}\label{VI-epsilon=1}
b=\sqrt{\frac ma}.
\end{equation}
Physicists  interpret $b$ as the length of the 
superconducting region in the 
normal material. This agrees with the behavior of the solution $u$
of Equation (\ref{VI-limeq}) which decays exponentially
at $-\infty$.\\
The boundary condition (\ref{VI-deGebndcn}) is derived by the
physicist de\,Gennes from the microscopic BCS theory. He considers  a planar
superconductor-normal junction in the absence of an applied magnetic
field (just as in Theorem~\ref{VI-sol-modeq}) and he
assumes 
firstly that no current passes through the boundary, and secondly
that there exists a  boundary condition of the form
$f(u,u_n,u_{nn},\dots)=0$; here the subscript $n$ denotes
differentiation in the normal direction of the boundary. What seems
interesting in our case is that we have determined the  boundary
condition (\ref{VI-deGebndcn}) in  the same situation of de\,Gennes,
but still in the framework of the Ginzburg-Landau (macroscopic)
theory.\\
For general domains, and in the regime $\varepsilon\to0$, we have
obtained in Theorem~\ref{VI-LuPaTh2}, just as predicted in the
physical literature (see ~\cite{deGe, deGe1}), a thin
superconducting sheath in $\Omega_2$ of thickness $\mathcal
O(\varepsilon)$. The `extrapolation length' now satisfies
\begin{equation}\label{VI-extrapolationlength}
b\approx\varepsilon\sqrt{\frac ma},
\end{equation}
hence it is decreasing with respect to $a$ and increasing with respect to $m$.
By the microscopic theory of superconductivity,
physicists are able to calculate both $a$ and $m$; one obtains actually that
$$a\approx T-T_c(\Omega_2),\quad m\approx \frac{\sigma_s}{\sigma_n}.$$
Here $T$ is the temperature, $T_c(\Omega_2)$ is the critical
temperature of the material in $\Omega_2$, $\sigma_s$ is the
conductivity of the superconducting material in ${\Omega_1}$ and
$\sigma_n$ that of the material in $\Omega_2$. Therefore, Formula
(\ref{VI-extrapolationlength}) shows that $b$
is both temperature and material dependent. Now the question that we
pose is about the dependence of $b$ on the
applied magnetic field $H$. According to \cite{deGeHu}, we expect
that $b$ is `essentially' field-independent when the intensity $H$
of the applied magnetic field is small, i.e. $H=o(1)$ as
$\varepsilon\to0$. However, when $H$ becomes of the order $\mathcal
O(1)$, we expect to observe a strong dependence of  $b$ on $H$.
Actually, we hope to prove that $b$ is a decreasing function of $H$.
This would prove the existence of the `breakdown field' $H_b$
predicted in the physical literature~\cite{Pa, deGeHu}. The field
$H_b$ is interpreted as the field at which it occurs the transition
from the Meissner state (phase of diamagnetic screening) to the
phase of magnetic field penetration in the normal material (i.e. in
${\Omega_2}$).

\subsection{Other asymptotic regimes}\ \\
It would be interesting to analyze  the asymptotic regimes $m\to+\infty$ or $m\to0_+$ (and this would also be physically relevant). Let us mention few remarks.
We look again at the solution $U$ (cf~.(\ref{VI-Sol})) of the equation in  $\mathbb R^2$.
%In this case, due to the symmetry, it is sufficient  (by scaling) to deal with the case $\varepsilon=1$, and hence the extrapolation length becomes~:
Let $u_N$ and $u_D$ be the positive bounded solutions of $-\Delta
u=(1-u^2)u$ in $\mathbb R\times\mathbb R_+$ with Neumann and
Dirichlet boundary conditions respectively. Then, as observed by
Lu-Pan in \cite{LuPa96} (see the remark p.~163 and Proposition~5.6), we have~:
$$u_N(x_1,x_2)=1,\quad u_D(x_1,x_2)=\frac{\beta\exp(\sqrt{2}x_2)-1}{\beta\exp(\sqrt{2}x_2)+1},\quad
\forall (x_1,x_2)\in\mathbb R\times\mathbb R_+.$$
Then it is readeable that~:
\begin{equation}\label{VI-m+inf,0+}
\lim_{m\to+\infty}\left\|U-u_N\right\|_{W^{1,\infty}(\overline{\mathbb R\times\mathbb R_+})}=0,
\quad\lim_{m\to0_+}\left\|U-u_D\right\|_{W^{1,\infty}(\overline{\mathbb R\times\mathbb R_+})}=0.
\end{equation}
Notice however, that in the regime $m\to+\infty$ the physical
interpretation of the extrapolation length $b$ in
(\ref{VI-extrapolationlength}) is no more accurate. %($b$ approximates
%the distance for which the superconducting Cooper electron pairs can
%penetrate into the normal
%material).\\
In view of (\ref{VI-m+inf,0+}) it seems reasonable to interpret 
Equations (\ref{VI-GL-LuPa}) (with $\gamma(\varepsilon)=0$) and
(\ref{VI-GL-D}) (with $g=0$) as limiting equations of (\ref{VIGL})
in the regimes $m\to+\infty$ and $m\to0_+$
respectively\footnote{Equation (\ref{VI-GL-D}) with $g=0$ is of
physical interest, since it is proposed in \cite{GP, HTW} as a model
for a superconductor  adjacent
 to a ferromagnetic material.}.\\
To make this rigorous, we denote by~:
$$C_N(\varepsilon):=\inf_{u\in H^1({\Omega_1})}\mathcal E(u),\quad
C_D(\varepsilon):=\inf_{u\in H^1_0({\Omega_1})}\mathcal E(u),$$
where the energy $\mathcal E$ is defined by~:
$$\mathcal E(u)=\int_{\Omega_1}\left(|\nabla u|^2+\frac1{2\varepsilon^2}(1-u^2)^2\right)\md x.$$
Furthermore, to emphasize the dependence on $m$, we write $C_0(\varepsilon,m)=C_0(\varepsilon)$, where
$C_0(\varepsilon)$ is introduced in (\ref{VI-infEn}). Then, it is clear that
\begin{equation}\label{VI-CN<Cm<CD}
C_N(\varepsilon)\leq C_0(\varepsilon,m)\leq C_D(\varepsilon).
\end{equation}
For large values of $m$, we get positive constants $C_\varepsilon$ and
$m_0$ such that,
\begin{equation}\label{VI-CN=Cm}
C_0(\varepsilon,m)\leq
C_N(\varepsilon)+\frac{C_\varepsilon}{\sqrt{m}},\quad\forall
\varepsilon>0,\quad\forall m\geq m_0.
\end{equation} 
To obtain (\ref{VI-CN=Cm}), it is sufficient to take
$\chi\left(\sqrt{m}\,t(x)\right)$ as a test function\footnote{We
recall that $C_N(\varepsilon)=\frac{|{\Omega_1}|}{2\varepsilon^2}$.}
(for the functional (\ref{VI-EnGL1})), where
$\chi$ is a cut-off satisfying (\ref{VI-chi}) and $t$ is the function in (\ref{VI-t(x)}).\\
In the regime $m\to0_+$, we believe that we shall have a lower bound of the following form~:
\begin{equation}\label{VI-CD=Cm}
C_D(\varepsilon)+\delta_\varepsilon(m)\leq C_0(\varepsilon,m).
\end{equation}
Here the function $\delta_\varepsilon$ satisfies $\displaystyle\lim_{m\to+\infty}
\delta_\varepsilon(m)=0$.\\
However, in the regime $m,\varepsilon\to0_+$ and $\frac{\sqrt{m}}\varepsilon\to\infty$,
we believe that our analysis  would permit us to obtain the following lower bound of the energy~:
\begin{equation}\label{VI-LB-Cm}
C_0(\varepsilon,m)\geq\left(\frac{2\sqrt2}3+o(1)\right)\frac{|\partial{\Omega_1}|}\varepsilon.
\end{equation}
%The proof follows  actually the same lines as in the Proof of Theorem~\ref{VIen}, Step~2.
We include here the additional constraint
$\frac{\sqrt{m}}{\varepsilon}\to+\infty$ in order to assure that the
use of Proposition~\ref{VI-int} is still possible.
%Notice also that the conclusion of Proposition~\ref{VI-bnd} (especially (\ref{VI-PtwConv})) is uniform with respect to $m$
%when $m$ remains in the bounded interval $]0,1]$. We must also mention that  we  use the second conclusion in (\ref{VI-m+inf,0+}).\\
Coming back to \cite{LuPa96}, it is proved that as $\varepsilon\to0_+$, we have,
\begin{equation}\label{VI-LP-D}
C_D(\varepsilon)=\left(\frac{2\sqrt2}3+o(1)\right)\frac{|\partial{\Omega_1}|}\varepsilon.
\end{equation}
Combined with (\ref{VI-LB-Cm}) and the upper bound in (\ref{VI-CN<Cm<CD}), one would obtain (in the regime
$m,\varepsilon\to0_+$ and $\frac{\sqrt{m}}\varepsilon\to+\infty$),
$$C_0(\varepsilon,m)=C_D(\varepsilon)(1+o(1)).$$
This explains why we expect that the lower bound (\ref{VI-CD=Cm}) is true.

\section*{Acknowledgements}
The author would like to thank B. Helffer for the interest he
owed to this work and for his many valuable
suggestions, and P. Sternberg for indicating the relevance of 
some references.\\
\indent This work has been supported by the European Research Network
  `Post-doctoral Training Program in Mathematical Analysis of Large
  Quantum Systems' with contract number HPRN-CT-2002-00277, the ESF
  Scientific Programme in Spectral Theory and Partial Differential
  Equations (SPECT), and the Agence Universitaire de la Francophonie (AUF).
  
\appendix

\section{Proof of Lemma~\ref{VIlem(4.1)}}\label{Appendix-Lemma}

This appendix is devoted to the proof of Lemma~\ref{VIlem(4.1)}, which
we state  again.
\begin{lem}\label{VIlem(4.1)-App}
Suppose  $u_\varepsilon\in C^2(\overline{\Omega_1})$ be a positive 
solution of $-\Delta
 u_\varepsilon=\frac1{\varepsilon^2}(1-u_\varepsilon^2)u_\varepsilon$ in $\Omega_1$. Then there exist positive constants 
$c_0,k_0,\varepsilon_0$ depending only on ${\Omega_1}$ such that,
\begin{equation}\label{VI-LuPa(4.3)-App}
\inf_{x\in{\Omega_1},t(x)\geq k_0\varepsilon}u_\varepsilon(x)\geq
c_0,\quad \forall~
\varepsilon\in]0,\varepsilon_0]. \end{equation}
Here $t$ is defined by (\ref{VI-t(x)}).\end{lem}
\paragraph{\bf Proof.}
We argue by contradiction. Assume that the conclusion of the lemma
were false. Then we may suppose (after passing to a subsequence) that
there is $x_\varepsilon\in \Omega_1$ such that, as $\varepsilon\to0$,
$$\frac{t(x_\varepsilon)}{\varepsilon}\to+\infty\quad{\rm and}\quad
u_\varepsilon(x_\varepsilon)\to0.$$
Let us define the following rescaled function
$$w_\varepsilon(x)=u_\varepsilon(x_\varepsilon+\varepsilon x)$$
on the rescaled domain 
$\Omega_\varepsilon=(\Omega_1-x_\varepsilon)/\varepsilon$.\\
Then, $w_\varepsilon$ satisfies the equation
$$-\Delta w_\varepsilon=(1-w_\varepsilon^2)w_\varepsilon\quad{\rm
  in}~\Omega_\varepsilon.$$
It is standard, as illustrated by the arguments in
  Section~\ref{VISec4}, to show that, after passing to a subsequence,
 $w_\varepsilon$ converges to some function $w$ in $C_{\rm
  loc}^2(\mathbb R^2)$, where $w$ solves the limiting equation:
$$-\Delta w=(1-w^2)w\quad{\rm in}\quad \mathbb R^2,$$
and satisfies the properties (infered from $u_\varepsilon$):
$$w(0)=0,\quad 0\leq w\leq 1.$$
Since $-\Delta w\geq0$ in $\mathbb R^2$, it results from the strong
maximum principle that $w\equiv0$ in $\mathbb R^2$. Having this point
in hand, we shall prove a contradiction.\\
Let us pick $R>0$ such that the first eigenvalue of the Dirichlet
Laplacian in $D(0,R)$ is equal to $1/3$. Let $\phi$ be the associated
positive and normalized eigenfunction,
$$-\Delta\phi=\frac13\phi,\quad \phi>0~{\rm in}~D(0,R),\quad
\phi=0~{\rm on}~\partial D(0,R).$$
Since $\|w_\varepsilon\|_{L^\infty(\Omega_\varepsilon)}\to0$ as
$\varepsilon\to0$, we get
$(1-w_\varepsilon^2)w_\varepsilon \geq\frac23w_\varepsilon$,  and
consequently one gets from the equation satisfied by $w_\varepsilon$,
$$\Delta w_\varepsilon+\frac23w_\varepsilon<0\quad{\rm in}~D(0,R).$$
Upon multiplying the above inequality by $\phi$ and integrating, and
since $w_\varepsilon,\phi>0$, we get
\begin{eqnarray*}
\int_{D(0,R)}\left(\phi\Delta w_\varepsilon-\phi\Delta
  w_\varepsilon\right)\,\md x\leq
\int_{D(0,R)}\left(\phi\Delta w_\varepsilon-\phi\Delta w_\varepsilon
  +\frac13\phi w_\varepsilon\right)\,\md x<0.\end{eqnarray*}
However, since the normal derivative $\partial\phi/\partial\nu<0$,
an integration by parts yields,
$$\int_{D(0,R)}\left(\phi\Delta w_\varepsilon-\phi\Delta
  w_\varepsilon\right)\,
\md x=-\int_{\partial D(0,R)}
  w_\varepsilon\frac{\partial\phi}{\partial\nu}\,\md s>0.$$
This is the requiered contradiction.\hfill$\Box$

\section{$L^2$-estimates for solutions of linear elliptic operators
with discontinuous coefficients} In this appendix we derive
$L^2$-estimates that permit us to prove
Theorem~\ref{VI-regKa}.\\
Let $\Omega_1,\Omega\subset\mathbb R^n$ ($n\geq1$) be open sets with
compact boundaries, and let
$\Omega_2=\Omega\setminus\overline\Omega_1$.\\
We consider the following linear elliptic operator
\begin{equation}\label{VI-L}
L=-{\rm div}\left(a(x)\nabla\right)+b(x)\cdot\nabla+c(x),
\end{equation}
where the coefficients $a,b,c$ are measurable functions in $\Omega$.
We suppose that the operator $L$ is uniformly elliptic, that is,
there exists $\lambda>0$ such that
\begin{equation}\label{VI-ellipticity}
a(x)\geq \lambda,\quad {\rm a.e.}\quad{\rm in }\quad \Omega.
\end{equation}
Given a function $f\in L^2(\Omega)$, we say that a function $u\in
H^1(\Omega)$ is a weak solution of
$$L\,u=f\quad{\rm in}\quad\Omega,$$
if the following condition holds
\begin{equation}\label{VI-appendix-weak}
\int_\Omega \left(a(x)\nabla u\cdot\nabla v
+(b(x)\cdot\nabla u)v +c(x)\,uv\right)\,\md x=\int_\Omega f\,v\,\md
x,\quad\forall v\in H_0^1(\Omega).
\end{equation}
Our objective is to prove the following theorem.
\begin{thm}\label{VI-appendix-th}
Suppose that the boundaries of $\Omega_1,\Omega$  are of class
$C^{k+2}$ ($k\geq 0$) and that the  coefficients satisfy
$$a\in C^{k+1}(\overline\Omega_1)\cup C^{k+1}(\overline\Omega_2),\quad
b,c\in C^{k}(\overline\Omega_1)\cup C^{k}(\overline\Omega_2).$$ There
exists a constant $C_k>0$ such that if
$u\in H^1_0(\Omega)$ is a solution of $Lu=f$ and if  $f$
satisfies
$$f_{|_{\Omega_1}}\in H^k(\Omega_1),\quad f_{|_{\Omega_2}}\in
H^k(\Omega_2),$$ then
\begin{equation}\label{VI-appendix-reg}
u_{|_{\Omega_1}}\in H^{k+2}(\Omega_1),\quad u_{|_{\Omega_2}}\in
H^{k+2}(\Omega_2),
\end{equation}
and we have the following estimate~:
\begin{equation}\label{VI-appendix-est}
\|u\|_{H^{k+2}(\Omega_1)}+\|u\|_{H^{k+2}(\Omega_2)}\leq
C_k\left(\|f\|_{H^k(\Omega_1)}+\|f\|_{H^k(\Omega_2)}+\|u\|_{L^2(\Omega)}\right).
\end{equation}
\end{thm}

The proof of Theorem~\ref{VI-appendix-th} is based on  the standard 
technique of difference quotients. Although many papers
are devoted to linear operators of the  type (\ref{VI-L}) (see
\cite{Tr} and references therein), Theorem~\ref{VI-appendix-th} is 
new. A natural (and
interesting) question is to ask for $L^p$ and H\"older type
estimates for solutions of linear PDE of the type (\ref{VI-L}).

\paragraph{\bf Proof of Theorem~\ref{VI-regKa}.}
It is sufficient to apply Theorem~\ref{VI-appendix-th} with $n=2$,
$$a(x)=\left\{\begin{array}{l}
1,\quad\text{in }\Omega_1\\
\displaystyle\frac1m,\quad\text{in }\Omega_2,
\end{array}\right.\quad
b\equiv0,\quad c\equiv0\quad\text{in }\Omega.
$$

\begin{lem}\label{VI-proof-k=0}
The conclusion of Theorem~\ref{VI-appendix-th} holds for $k=0$.
Moreover, the solution $u$ satisfies the following boundary
condition on $\partial\Omega$~:
\begin{equation}\label{VI-appendix-BC}
\mathcal T_{\partial\Omega_1}^{\rm int}(a(x)\,\nu\cdot\nabla
u)=\mathcal T_{\partial\Omega_1}^{\rm ext}(a(x)\,\nu\cdot\nabla u).
\end{equation}
\end{lem}
\paragraph{\bf Proof.}
Let $\chi\in C^2_0(\mathbb R^n)$ be a cut-off  function  with
support in a ball $B_R$
centered at a point on $\partial\Omega$ and such that
$$0\leq\chi\leq1,\quad \chi\equiv1\quad\text{ in }B_{R/2}.$$
By standard regularity
theory, it will be sufficient to prove that $$\chi u\in
H^2(\Omega_1)\cup H^2(\Omega_2).$$ 
Since the boundary of $\Omega_1$ is smooth of class $C^2$, we shall
work in a coordinate system $\widetilde x=(\widetilde x_1,\widetilde
x_2,\dots,\widetilde x_n)$ such that the boundary of $\Omega_1$ in
the support of $\chi$ is defined by $\{\widetilde x_n=0\}$,
 and the transformation $x\mapsto\widetilde x$ is of
class $C^2$. We remark also that $\Omega_1$ and $\Omega_2$ are
defined now by $\{\widetilde x_n>0\}$ and $\{\widetilde x_n<0\}$
respectively. To a function $v$ defined in the $x$-coordinate
system, we assign a function $\widetilde v$ defined in the
$\widetilde x$-system by
$\widetilde v(\widetilde x)=v(x)$.\\
The weak formulation (\ref{VI-appendix-weak}) becomes now~:
\begin{equation}\label{VI-appendix-weak'}
\int_{\mathbb R^n}\left\{\widetilde a(\widetilde
x)\widetilde\nabla_{\widetilde x}\widetilde
u\cdot\widetilde\nabla_{\widetilde x}\widetilde v+(\widetilde
b(\widetilde x)\cdot\widetilde\nabla_{\widetilde x}\widetilde
u)\widetilde v+\widetilde c(\widetilde x)\,\widetilde u\,\widetilde
v(x)\right\}J\,\md\widetilde x=\int_{\mathbb R^n}\widetilde
f\,\widetilde v\,J\,\md\widetilde x. \end{equation} Here $J$ is the
Jacobian of the
transformation $x\mapsto\widetilde x$.\\
For $1\leq j\leq n-1$, we define the following test function~:
\begin{equation}\label{VI-appendix-testfunction}
\widetilde v(\widetilde
x)=D_{j,-h}\left(D_{j,h}\widetilde\chi^2 \widetilde
u\right)(\widetilde x).
\end{equation} Here, the {\it difference
quotient} $D_{j,h}$ is defined by
$$D_{j,h}u(\widetilde x)=\frac1h\left[u(\widetilde
x+he_j)-u(\widetilde x)\right],$$ where $\{e_j\}_{j=1}^n$ is the
canonical
orthonormal basis of $\mathbb R^n$.\\
Substituting the test function $\widetilde v$ in
(\ref{VI-appendix-weak'}), we get (we remove the tildas for
simplicity of notation)~:
\begin{eqnarray}\label{VI-appendix-weak''}
&&\int_{\mathbb R^n}\left\{[D_{j,h}(a\nabla u)]\cdot
[D_{j,h}(\chi^2\nabla u)]+(b\cdot \nabla
u)(D_{j,-h}[D_{j,h}(\chi^2u)])\right.\\
&&\hskip2cm\left.+c\,(D_{j,h}u)(D_{j,h}\chi^2u)\right\}\,J\,\md
x=\int_{\mathbb R^n}f[D_{j,-h}(D_{j,h}\chi^2u)]\,J\,\md x.\nonumber
\end{eqnarray}
By the ellipticity condition (\ref{VI-ellipticity}) and the
hypothesis on $a$, and since $j\not=n$, we get positive constants
$C_1,C_2$ and $h_0$ such that we have for $h\in]0,h_0]$~:
\begin{eqnarray}\label{VI-appendix-1}
&&\int_{\mathbb R^n}[D_{j,h}(a\nabla u)]\cdot [D_{j,h}(\chi^2\nabla
u)]\,J\,\md x\\
&&\geq C_1\int_{\mathbb R^n}|D_{j,h}(\chi \nabla
u)|^2\,J\,\md x
-C\int_{\mathbb R^n}(|\chi|^2+|\nabla\chi|^2)|\nabla u|^2\,J\,\md x.\nonumber
\end{eqnarray}
We mention here that the constants $C_1$ and $C_2$ are controlled by
$\|a\|_{C^1(\overline\Omega_{i})}$ ($i=1,2$),
$\|\chi\|_{C^1(\overline\Omega)}$  and  the ellipticity
constant $\lambda$.\\
We get also, by applying the Cauchy-Schwarz Inequality~:
\begin{eqnarray}\label{VI-appendix-2}
&&\left|\int_{\mathbb R^n} (b\cdot \nabla
u)(D_{j,-h}[D_{j,h}(\chi^2u)])\,J\,\md x\right|
\\
&&\leq \frac{C_1}4\int_{\mathbb
R^n}|D_{j,h}(\chi\nabla u)|^2\,\md x+C_2\int_{\mathbb R^n}(|\chi|^2+|\nabla\chi|^2)|\nabla u|^2\,J\,\md
x,
\end{eqnarray}
where the constant $C_2>0$ is controlled by
$\|b\|_{L^\infty(\Omega)}$ and the constant $C_1$ introduced in (\ref{VI-appendix-1}).\\
We get also by the Cauchy-Schwarz Inequality~:
\begin{eqnarray*}\label{VI-appendix-3}
&&\left|\int_{\mathbb R^n}f[D_{j,-h}(D_{j,h}\chi^2u)]\,J\,\md
  x\right|\\
&&\leq
\frac{C_1}4\int_{\mathbb R^n}|D_{j,h}(\chi \nabla u)|^2\,J\,\md
x+C\int_{\mathbb R^2}\left(|\chi f|^2+(|\chi|^2+|\nabla\chi|^2)\right)|\nabla
u|^2\,J\,\md x.\nonumber
\end{eqnarray*}
Substituting (\ref{VI-appendix-1})-(\ref{VI-appendix-3}) in
(\ref{VI-appendix-weak''}), we get~:
$$\int_{\mathbb R^n}|D_{j,h}(\chi u)|^2\,J\,\md x\leq C\int_{\mathbb
R^n}\left(|\chi f|^2+|\chi \nabla u|^2\right)\,J\,\md x,$$ for a
constant $C>0$ and for $1\leq j\leq n-1$. As  $h\to0$, we get
that 
$$\partial_{x_j}\nabla (\chi u)\in L^2(\mathbb R^n),\quad\text{ for } 1\leq
j\leq n.$$ So what remains to prove is that $[\partial_{x_n}^2(\chi
u)]_{|_{\mathbb R^{n-1}\times\mathbb R_\pm}}\in L^2(\mathbb
R^{n-1}\times\mathbb R_\pm)$. This is actually evident since, by
coming back to the equation satisfied by $u$, it reads now in the
form~:
$$\partial_{x_n}(a(x)\partial_{x_n})(\chi u)\in L^2(\mathbb R^n).$$
Since $a\in C^1(\overline{\mathbb R^{n-1}\times\mathbb R_\pm})$ (and
hence it may have singularities  through $\{x_n=0\}$), this would only
give that $[\partial_{x_n}^2(\chi u)]_{|_{\mathbb
R^{n-1}\times\mathbb R_\pm}}\in L^2(\mathbb R^{n-1}\times\mathbb
R_\pm)$. This achieves now the proof of the
Lemma.\hfill$\Box$\\

\paragraph{\bf Proof of Theorem~\ref{VI-appendix-th}.}
The proof is by induction on $k$. Assume that the result of the
theorem holds for a given integer $k\geq 0$. We have then to
establish~:
$$f_{|_{\Omega_{i}}}\in H^{k+1}(\Omega_i)\Rightarrow
u_{|_{\Omega_{i}}}\in H^{k+3}(\Omega_i)\quad (i=1,2).$$ By the
induction hypothesis, we already have $u_{|_{\Omega_i}}\in
H^{k+2}(\Omega_i)$, and by Lemma~\ref{VI-proof-k=0}, $u$ satisfies the
boundary condition (\ref{VI-appendix-BC}).\\
We work again in the coordinates system $\widetilde x$ introduced in
the proof of
Lemma~\ref{VI-proof-k=0}, and we remove the tildas in order to simplify notations.\\
Let us consider again a cut-off function
$\chi^k\in C_0(\mathbb R^n)$  with support in a ball
centered at a point on $\partial\Omega$. We assume that $\chi^k$ is  continuous,
$$(\chi^k )_{|_{{\Omega_i}}}\in C^{k+2}(\overline{\Omega_i}),\quad
 i=1,2,$$
and $\chi^k $ satisfies the boundary condition
\begin{equation}\label{VI-appendix-BC-chi}
\mathcal T_{\partial\Omega_1}^{\rm int}(a(x)\,\nu\cdot\nabla
\chi^k )=\mathcal T_{\partial\Omega_1}^{\rm ext}(a(x)\,\nu\cdot\nabla
\chi^k ).
\end{equation}
One can indeed select $\chi^k $ in the following way (recall that we
work in a coordinate system such that the boundary of $\Omega_1$ is
given by $\{x_n=0\}$)~:
$$\mathbb R^{n-1}\times\mathbb R\ni(x',x_n)\mapsto\chi^k (x',x_n)=\left\{
\begin{array}{l}
\psi(x')\,\widetilde\chi(x_n),\quad \text{if }x_n>0,\\
\psi(x')\,\widetilde\chi\left(\displaystyle\frac{a(x',0_-)}{a(x',0_+)}x_n\right),\quad\text{if
}x_n\leq0.
\end{array}\right.
$$
Here $\psi$ and $\widetilde \chi$ are  cut-off functions such that
$\psi$ is supported in a ball of $\mathbb R^{ n-1}$ (centered at the
origin and of sufficiently small radius $t_0\in]0,1[$)  and
$\widetilde\chi$ satisfies~:
$$\widetilde \chi\in C^\infty(\mathbb R),\quad0\leq\widetilde \chi\leq1,\quad\widetilde
\chi\equiv1\quad\text{ in }[-t_0/2,t_0/2],\quad{\rm
  supp}\,\widetilde\chi\subset[-t_0,t_0].$$  
By standard regularity theory, it is sufficient to prove that $\chi^k 
u\in H^{k+3}(\Omega_i)$ $(i=1,2)$.\\
For $1\leq j\leq n-1$, we introduce the following test function~:
$$v=\chi^k  D_{j,h}u.$$
Since
$\chi^k $ satisfies the boundary condition (\ref{VI-appendix-BC-chi}),
we get through the application of Lemma~\ref{VI-proof-k=0} that $v$
satisfies also the boundary condition in (\ref{VI-appendix-BC}).\\
Therefore, $v$ is a weak solution of  an elliptic equation of the form~:
$$\widetilde L\,v=\widetilde f,\quad{\rm in}\quad\mathbb R^n,$$
where $\widetilde L$ is an operator of the type (\ref{VI-L}), and
$\widetilde f\in H^k(\mathbb R^{n-1}\times\mathbb R_\pm)$. 
Now, we have by the induction hypothesis, \begin{eqnarray*} &&\|\chi^k 
D_{j,h} u\|_{H^{k+2}(\mathbb R^{n-1}\times\mathbb R_\pm)}\\
&&\leq C\left(\|\widetilde f\|_{H^{k}(\mathbb R^{n-1}\times\mathbb
R_+)}+\|\widetilde f\|_{H^k(\mathbb R^{n-1}\times\mathbb R_-)}+\|\chi^k 
D_{j,h}u\|_{L^2(\mathbb R^n)}\right),\end{eqnarray*} where
$\|\widetilde f\|_{H^k(\mathbb R^{n-1}\times\mathbb R_\pm)}$ is
controlled by $\|f\|_{H^{k+1}(\Omega_i)}$ and
$\|u\|_{H^{k+1}(\Omega_i)}$ $(i=1,2)$. Therefore, upon making
$h\to0$, we get that 
$$\chi^k \partial_{x_j}u\in H^{k+2}(\mathbb
R^{n-1}\times\mathbb R_\pm),\quad\text{ for }1\leq j\leq n-1.$$ We get also from
the equation satisfied by $u$ that $\chi^k \partial_{x_n}u\in
H^{k+2}(\mathbb R^{n-1}\times\mathbb R_\pm)$ which establishes the
theorem up to the order $k+1$.\hfill$\Box$

\end{document}